\documentclass[journal]{IEEEtran}
\usepackage{lineno,hyperref}
\modulolinenumbers[5]
\usepackage{algorithmic}
\usepackage{graphicx}
\DeclareGraphicsExtensions{.pdf,.jpeg,.png}
\usepackage{subfigure}
\usepackage{booktabs}
\usepackage{multirow}
\usepackage{array}
\usepackage{caption,setspace}
\usepackage{graphics}
\usepackage{makecell}
\usepackage[table]{xcolor}

\ifCLASSINFOpdf
 
\else
\fi

\begin{document}
%
\title{Separable Reversible Data Hiding Based on Integer Mapping and Multi-MSB Prediction for Encrypted 3D Mesh Models}
%
%
%

\author{Zhaoxia~Yin,~\IEEEmembership{Member,~IEEE,}~Na~Xu and ~Feng~Wang
\thanks{This research work is partly supported by National Natural Science Foundation of China (61872003, 61860206004).}
\thanks{Zhaoxia Yin and Na Xu are with the school of Computer Science and Technology, Anhui University, e-mail: yinzhaoxia@ahu.edu.cn.}
}

\markboth{IEEE Transactions on Multimedia}
{Shell \MakeLowercase{\textit{et al.}}: } 
\maketitle

\maketitle

\begin{abstract}
Reversible data hiding in encrypted domain (RDH-ED) has received tremendous attention from the research community because data can be embedded into cover media without exposing it to the third party data hider and the cover media can be losslessly recovered after the extraction of the embedded data. Although, in recent years, extensive studies have been carried out about images based RDH-ED, little attention is paid to RDH-ED in 3D meshes due to its complex data structure and irregular geometry. In this paper, we propose a separable RDH-ED method for 3D meshes based on integer mapping and Multi-MSB (multiplication most significant bit) prediction.  
The proposed method divides all the vertices of the mesh into the ``embedded" set and ``reference" set, and maps decimals of the vertex into integers. Then, we calculate the Multi-MSB prediction errors for the vertices of the ``embedded" set and a bit-stream encryption technique will be executed. Finally, additional data is embedded by replacing the Multi-MSB of the encrypted vertex coordinates. According to different permissions, recipient can obtain the original plaintext meshes, additional data or both. Experimental results show that the proposed method has higher embedding capacity and higher quality of the recovered meshes compared to the state-of-art methods.
\end{abstract}

\IEEEpeerreviewmaketitle

\section{Introduction}

Reversible data hiding (RDH) allows to embed additional data into cover media in a way that the cover media can be losslessly restored after the extraction of embedded data. In recent years, RDH has been extensively studied by the research community~\cite{ref_article1,ref_article2,ref_article3,ref_article4,ref_article5,ref_article6,ref_article7,ref_article8}.
The existing RDH schemes can be mainly divided into three categories: lossless compression, difference expansion(DE) and histogram shifting(HS). Image lossless compression~\cite{ref_article1,ref_article2} extracts some features of the original image for lossless compression, and additional data is embedded into the reserved space left by lossless compression. Lossless compression based RDH schemes offer low embedding capacity. To improve the embedding capacity, difference expansion (DE)~\cite{ref_article3,ref_article4,ref_article5} and histogram shifting (HS)~\cite{ref_article6,ref_article7,ref_article8} based RDH methods emerged. 

With third party cloud paradigm and privacy preserving applications, the demand for privacy protection is increasing. The combination of encryption and RDH technologies play a crucial role in privacy protection. In order to store or share files securely using third party services, the content owner will use encryption technology to transform the original content into unreadable ciphertext before transmission, and then data hider will embed data into ciphertext for data management, authentication and ownership protection. At the same time, recipient want to be able to recover the original content losslessly after decryption and data extraction. Such privacy preserving scenarios trigger RDH-ED for managing ciphertext data.

In recent years, some existing RDH-ED algorithms for embedding additional data into encrypted carriers have been proposed, which can be classified into two categories: vacating room after encryption (VRAE)~\cite{ref_article9,ref_article10,ref_article11} and reserving room before encryption (RRBE)~\cite{ref_article12,ref_article13}.
Zhang~\cite{ref_article9} first proposed the RDH-ED method, which embedded additional data by modifying the pixels of encrypted images and extracted data by using pixel correlation of original images. The original encrypted image is divided into non-overlapping blocks and embedded data by flipping the three least significant bits (LSB) of half of the pixels in each block. Due to the spatial correlation of the image, the original image block is much smoother than the modified image block, so the recipient used a smoothness estimation function to estimate the texture complexity of each block for data extraction and image restoration. However, the quality of restored image and the accuracy of data extraction are still not very satisfactory.
\begin{figure}[!ht]
  \centering
\vspace{-0.2cm}

\setlength{\belowcaptionskip}{-0.75cm}
   \includegraphics[width=0.5\textwidth]{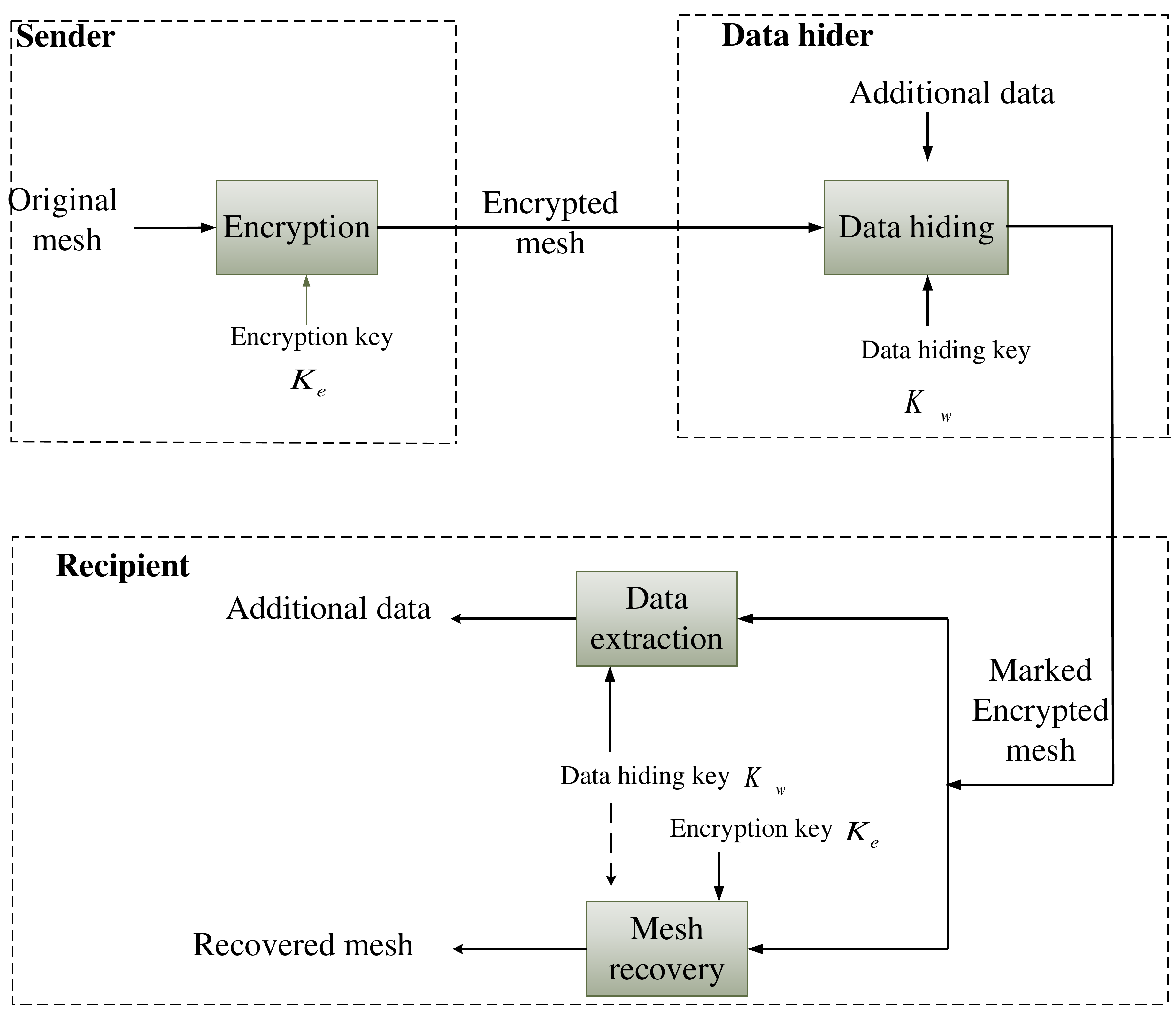}
 \caption{The framework of RDHED methods based on 3D mesh models.}
\label{fig1}
\end{figure}
\vspace{-0.03cm}
In the method mentioned above, data extraction and images recovery are carried out at the same time. In order to separate data extraction from image recovery, Zhang~\cite{ref_article10} proposed a separable RDH-ED scheme based on LSB compression. This scheme realizes that additional data can be extracted directly from embedding room and images can be restored without data extraction. After that, Qian et al.~\cite{ref_article11} proposed a scheme that uses stream cipher to encrypt the original images. The data-hider compresses a series of selected bits taken from the encrypted images to make room for additional data. If the recipient has the embedding key and the encryption key, then the additional data can be extracted correctly and original images can be perfectly restored using distributed source coding.

Compared with VRAE algorithms, RRBE algorithms have better performance in reducing data extraction errors and restoring original images. 
Ma et al.~\cite{ref_article12} proposed to reserve room for data embedding before encryption. Ma et al.'s method can guarantee that there are no errors in data extraction and image recovery. Most recently, Puteaux et al.~\cite{ref_article13} proposed a RDH-ED scheme in which data hider embeds additional data into the pixels of encrypted images through MSB substitution. The recipient can extract additional data from the MSB plane of the encrypted images. After decryption, the recipient used correlation between adjacent pixels, and reconstructs the original images through MSB prediction.

According to the the atart-of-the-art methods introduced above, RDH algorithms for images have been extensively studied for many years, but these algorithms cannot be directly applied to other cover media, such as text, audio, video, and 3D models. Because of the wide range of applications of 3D models (mesh models or point cloud models) and its large intrinsic capacity, 3D models are considered potential cover carriers for RDH. However, research on RDH technology using 3D models as cover media is still in initial stage. Therefore, the study of 3D models is a very valuable topic. According to literature, existing RDH methods of 3D models are mainly divided into four domains: spatial domain, transform domain, compressed domain and encrypted domain.

Spatial domain RDH methods~\cite{ref_book1,ref_article14,ref_article15} embedded additional data into 3D models by slightly modifying vertex coordinates. Transform domain RDH methods~\cite{ref_book2} embedded additional data into the transformation coefficients of the models. Compressed domain RDH methods~\cite{ref_book3,ref_article16} use vector quantization(VQ) for compressing the vertices of 3D models and then embedded data into compressed meodels stream. 
To our best of knowledge, there are only two papers on 3D meshes based RDH-ED~\cite{ref_article17}~\cite{ref_article18}. Fig.1 shouws the framework of RDHED methods  using 3D models as cover media precisely. Jiang et al.~\cite{ref_article17} mapped vertex coordinates of 3D meshes to integers using scaling and quantization and then obtain the encrypted meshes. Additional data is embedded by flipping several LSBs of the encrypted coordinates. At the recipient side, the marked encrypted mesh is firstly decrypted and then realizes data extraction and mesh reconstruction by using the designed smoothing measure function. Jiang et al.~\cite{ref_article17} is an inseparable scheme, i.e. meshes decryption and data extraction are carried out at the same time.~\cite{ref_article17} has the disadvantages of low embedding capacity, poor reconstruction meshes quality and larger errors in data extraction. In~\cite{ref_article18}, a two-tier RDH-ED method for 3D meshes using homomorphic Paillier cryptosystem is proposed, which is suitable for cloud data management. Due to the large ciphertext expansion and high computational complexity of the Paillier cryptosystem, ~\cite{ref_article18} is not efficient in practice. 

In this paper, We took disadvantages of~\cite{ref_article17}~\cite{ref_article18} into account and proposed a separable reversible data hiding based on integer mapping and multi-MSB prediction for encrypted 3D meshes. Our method first maps signed floating vertex coordinates values to integer values, analyzes and finds out the vertex index information with prediction error in the original 3D mesh as auxiliary information, and then uses stream cipher to encrypt the original 3D mesh. According to different permissions, recipient can obtain the original plaintext mesh, additional data or both. Compared with state-of-the-art methods, we improves the embedding capacity and the quality of  recovered mesh, and realizes error-free extraction of the data.

The main contributions of this paper are as follows:

 1) The Multi-MSB embedding strategy was adopted to improve the embedding capacity.

2) The proposed method can obtain high quality recovered mesh by ring -prediction, and the data extraction and mesh recovery are separable and error- free.

The rest of this paper is organized as follows: Section II introduces  the proposed method. Section III presents the analysis of experimental results. Section IV concludes this paper and descibe the future work.
\begin{figure*}[h]
\centering
 \vspace{-0.4cm} 
\setlength{\belowcaptionskip}{-0.5cm}
    \includegraphics[width=\textwidth]{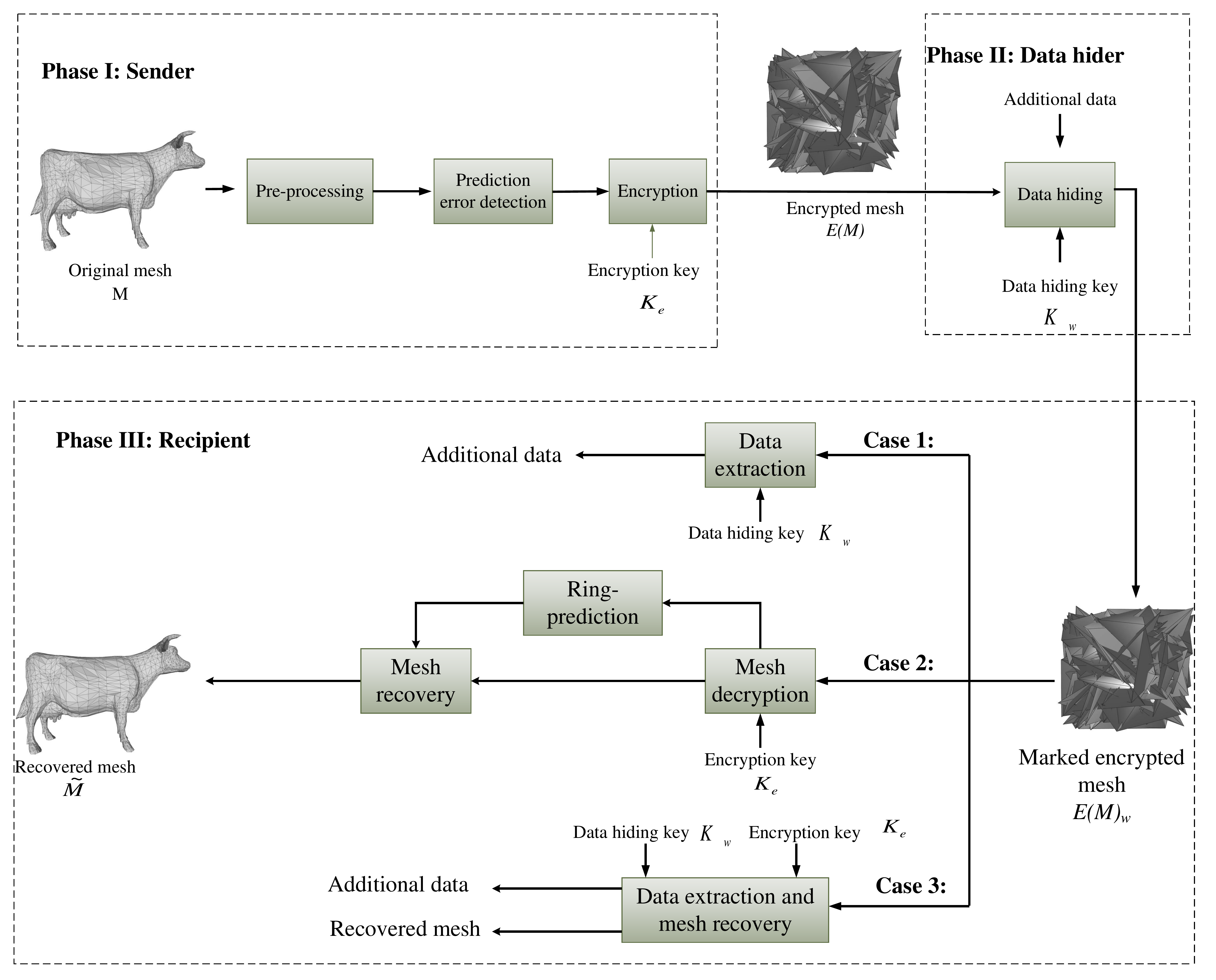}
  \caption{Framework of the Proposed Method.}
\label{fig2}
\end{figure*}


 

\section{Proposed method}

In this paper, our goal is to hide additional data in 3D mesh by slightly modifying the mesh vertices without changing the mesh topology, while allowing the recipient to correctly extract additional data and perfectly recover the original mesh. 
The method consists of pre-processing, prediction error detection, encryption, data hiding, data extraction and mesh recovery. Fig. 2 illustrates  framework of the proposed method.

First, the sender analyzes the original 3D mesh to find the vertex index number information with prediction error and records it as auxiliary information. Then, the sender uses the encryption key \emph{Ke} to encrypt the original mesh ${M}$ to get the encrypted 3D mesh ${E(M)}$. In the data embedding stage, additional data is embedded into the encrypted mesh ${E(M)}$ to obtain the marked encrypted mesh ${E(M)w)}$. After the mesh encryption and data embedding stage, the sender sends the encrypted auxiliary information along with ${E(M)w}$ as supplementary data.
There are three cases for the recipient. Case 1: For the recipient has only the data hiding key ${Kw}$, the recipient can extract additional data from ${E(M)w)}$. Case 2: For the recipient has only the encryption key ${Ke}$, a high-quality recovered mesh  $\widetilde{M}$ can be obtained. Case 3: For the recipient with encryption ${Ke}$ and data hiding keys ${Kw}$, the recipient can extract the data and recover the original mesh.
\vspace{-0.5cm}
\subsection{Pre-processing}
3D mesh models are represented in various file formats such as OFF, PLY, OBJ, etc. The 3D mesh is composed of vertices data and faces data. Vertices data include coordinates data of vertices represented as \emph{V}= \{\emph{v$_i$} $\in$ $\Re$$^3$$\vert$1$\le$\emph{i}$\leq$\emph{N}\} , where the vertex is represented as \emph{v$_{i}$}=(\emph{v$_{i,x}$},\emph{v$_{i,y}$},\emph{v$_{i,z}$}), and \emph{N} is the number of vertices. Note that each coordinate \emph{v$_{i,j}$} < 1 and \emph{j}$\in$$\{$\emph{x,y,z}$\}$. \emph{F}=(\emph{f$_1$},\emph{f$_2$}\ldots\emph{f$_\emph{M}$}) represent faces sequence when we traverse the faces data, where \emph{f$_{i}$}=(\emph{v$_{i,x}$},\emph{v$_{i,y}$},\emph{v$_{i,z}$}), ${M}$ is the number of faces. We take a local region of a “Cow” mesh for illustration in Fig. 3, and the corresponding file format is shown in Table I. 
\begin{figure*}[!ht]
\vspace{-0.4cm} 
\setlength{\belowcaptionskip}{-0.25cm}
  \centering
    \includegraphics[width=0.9\textwidth]{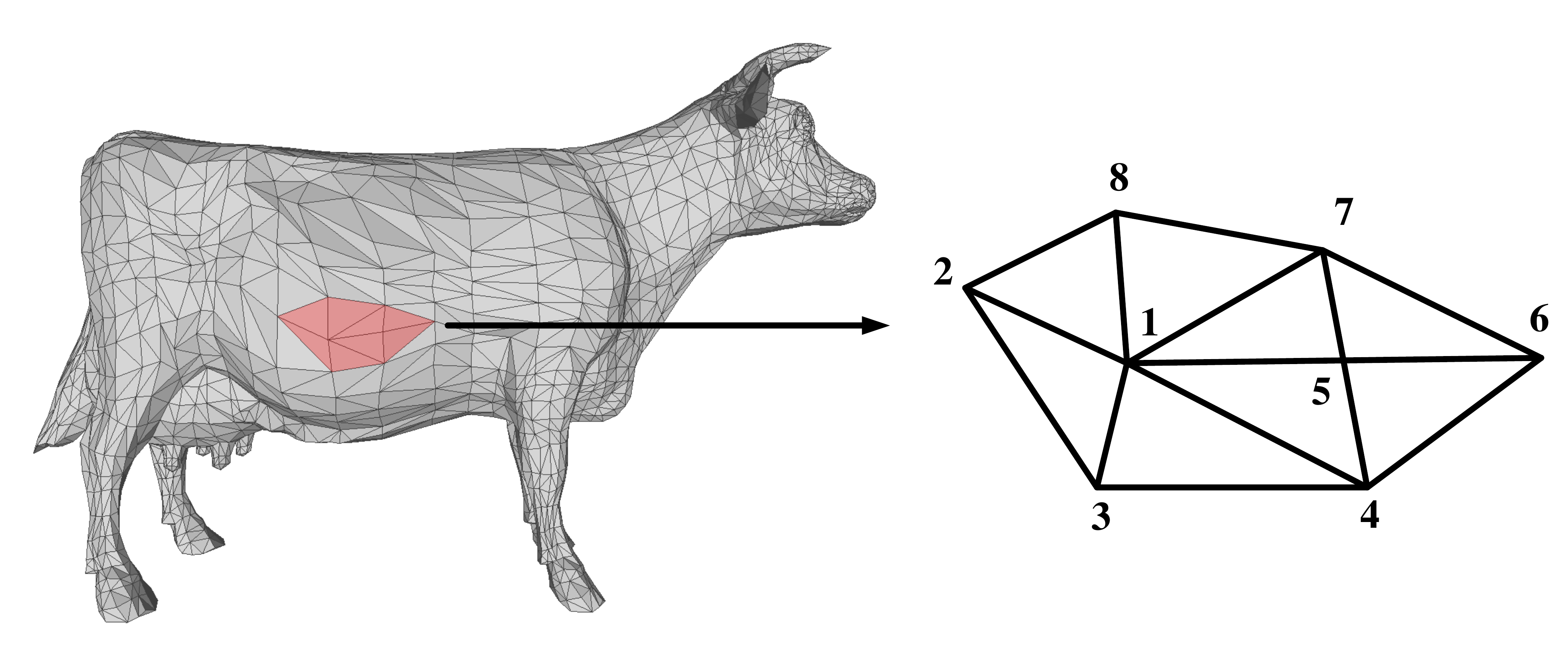}
  \caption{Cow Mesh.}
\label{fig3}
\end{figure*}

\begin{table*}[!ht]
\begin{center}
\setlength{\tabcolsep}{3mm}
\setlength{\belowcaptionskip}{-0.15cm} 
\newcommand{\tabincell}[2]{\begin{tabular}{@{}#1@{}}#2\end{tabular}}
\caption{\label{tb::tab1} File format for Fig. 2.}
\begin{tabular}{|c|c|c|c|c|c|c|}
\hline
{\tabincell{l}{Index of \\ vertex}} &\tabincell{l}{x-axis\\ coordinate} & \tabincell{l}{y-axis  \\coordinate} & \tabincell{l}{z-axis \\coordinate} & \tabincell{l}{coordinates}& \tabincell{l}{Index\\of face} & \tabincell{l}{Elements in\\each face  \\} \\
\hline
1 & \emph{v$_{1,x}$} & \emph{v$_{1,y}$}  &  \emph{v$_{1,z}$} & (0.180757, 0.034214, 0.193897)   & 1  & (1,2,8) \\
2 &  \emph{v$_{2,x}$} & \emph{v$_{2,y}$} &  \emph{v$_{2,z}$}&(0.118210, 0.059086, 0.189724)    &2   & (7,8,1)\\
3 & \emph{v$_{3,x}$} & \emph{v$_{3,y}$} &  \emph{v$_{3,z}$} &(0.092150, 0.029539, 0.197267)    & 3  & (5,7,1)\\
4 & \emph{v$_{4,x}$} & \emph{v$_{4,y}$} &  \emph{v$_{4,z}$} &(0.137215, 0.043615, 0.201492)    & 4  & (1,5,4)\\
5 & \emph{v$_{5,x}$} & \emph{v$_{5,y}$} &  \emph{v$_{5,z}$} &(0.136288, 0.065522, 0.187564)    & 5  & (3,4,1)\\
6 & \emph{v$_{6,x}$} & \emph{v$_{6,y}$} &  \emph{v$_{6,z}$} &(0.160892, 0.015154, 0.200969)    & 6  & (2,1,3)\\
\ldots & \ldots &\ldots & \ldots & \ldots & \ldots& \ldots\\
20& \emph{v$_{20,x}$} & \emph{v$_{20,y}$} &  \emph{v$_{20,z}$}  &(0.082017, 0.026986, 0.253443)& 20 & (20,19,23)\\
21& \emph{v$_{21,x}$} & \emph{v$_{21,y}$} &  \emph{v$_{21,z}$}  &(0.026661, 0.037672, 0.246828)& 21 & (21,20,34)\\
\ldots & \ldots &\ldots & \ldots & \ldots & \ldots & \ldots\\
\hline
\end{tabular}
\end{center}
\end{table*}

We can perform lossy compression of vertex coordinates according to the recommendation of~\cite{ref_article19}. According to the different precision ${m}$, the corresponding integer value is between 0 and $2^\emph{${m}$}$, where \emph{${m}$}$\in$[1-33].
Normalizing floating point coordinates \emph{v$_{i,j}$} to integer coordinates \emph{$\bar{v}$$_{i,j}$} as
\begin{equation}
\emph{$\bar{v}$$_{i,j}$}=\lfloor \emph{v$_{i,j}$}\times  10^\emph{${m}$}\rfloor,
\end{equation}
Where \emph{${i}$} is the ith vertex, \emph{j}$\in$$\{$\emph{x,y,z}$\}$, \emph{v$_{i,j}$} is the original set of floating point vertices and \emph{$\bar{v}$$_{i,j}$} is the set of integer vertices.

Recipient can convert the processed integer coordinates to floating point coordinates by Eq. (2).
\begin{equation}
\emph{$\hat{v}$$_{i,j}$}=\emph{$\bar{v}$$_{i,j}$}  / 10^\emph{${m}$},
\end{equation}
\emph{${m}$} is an important parameter for determining the lossless restory recovery of the mesh. The value of \emph{${m}$} corresponds to the bit-length \emph{${l}$} of integer coordinates as 
\begin{equation}
\emph{l} = \left\{ \begin{array}{ll}
8,\qquad\quad &\textrm{1$\le$\emph{${m}$} $\leq$ 2}\\
16,\qquad \quad& \textrm{3$\le$\emph{${m}$} $\leq$ 4}\\
32,\qquad \quad& \textrm{5$\le$\emph{${m}$} $\leq$ 9}\\
64,\qquad\quad & \textrm{10$\le$\emph{${m}$} $\leq$ 33},
\end{array} \right.
\end{equation}
Different \emph{${m}$} correspond to different bitlength \emph{${l}$}, which means that \emph{${m}$} affect the quality of the recovery mesh and the time cost of each stage, including encryption, data embedding, data extraction and mesh recovery.
For example, for floating point vertex \emph{v} = (-0.202018, -0.0740184, 0.288808) and \emph{m} = 4, the corresponding integer coordinates are \emph{$\bar{v}$} = (-2020, -7400, 2888). The integer range of \emph{$\bar{v}$} is 0-9999, and the required storage bit length is 16.
\subsection{Prediction Error Detection}
The ``embedded" set \emph{C} is used to embed additional data, and the ``reference" set \emph{R} is used to recover the mesh without modifying the vertices during the whole process.
The sender traverses all the vertices in faces data in ascending order, and calculate ``embedded" set \emph{C} and ``reference" set \emph{R} according to topological information between vertices. Sender traverses the first vertex in the face data and adds this vertex to \emph{C}, find its adjacent vertices and add them to \emph{R}. 
As shown in Fig. 3, when traversing the Cow's faces sequence, sender first traverses to a vertex numbered 1, and add this vertex to \emph{C}. Then, sender traverses the faces sequence to find the faces containing 1, and in this way sender find 2, 3, 4, 5, 7, 8 as its adjacent vertices, and then all the adjacent vertices numbered 2, 3, 4, 5, 7, 8 are added to \emph{R}.
Finally, we test whether the vertices in the \emph{C} have prediction error and the maximum embedding length of the embeddable vertices. The vertices with prediction errors are not used for data embedding.
\begin{figure*}[!th]
\begin{center}  
\includegraphics[width=\textwidth]{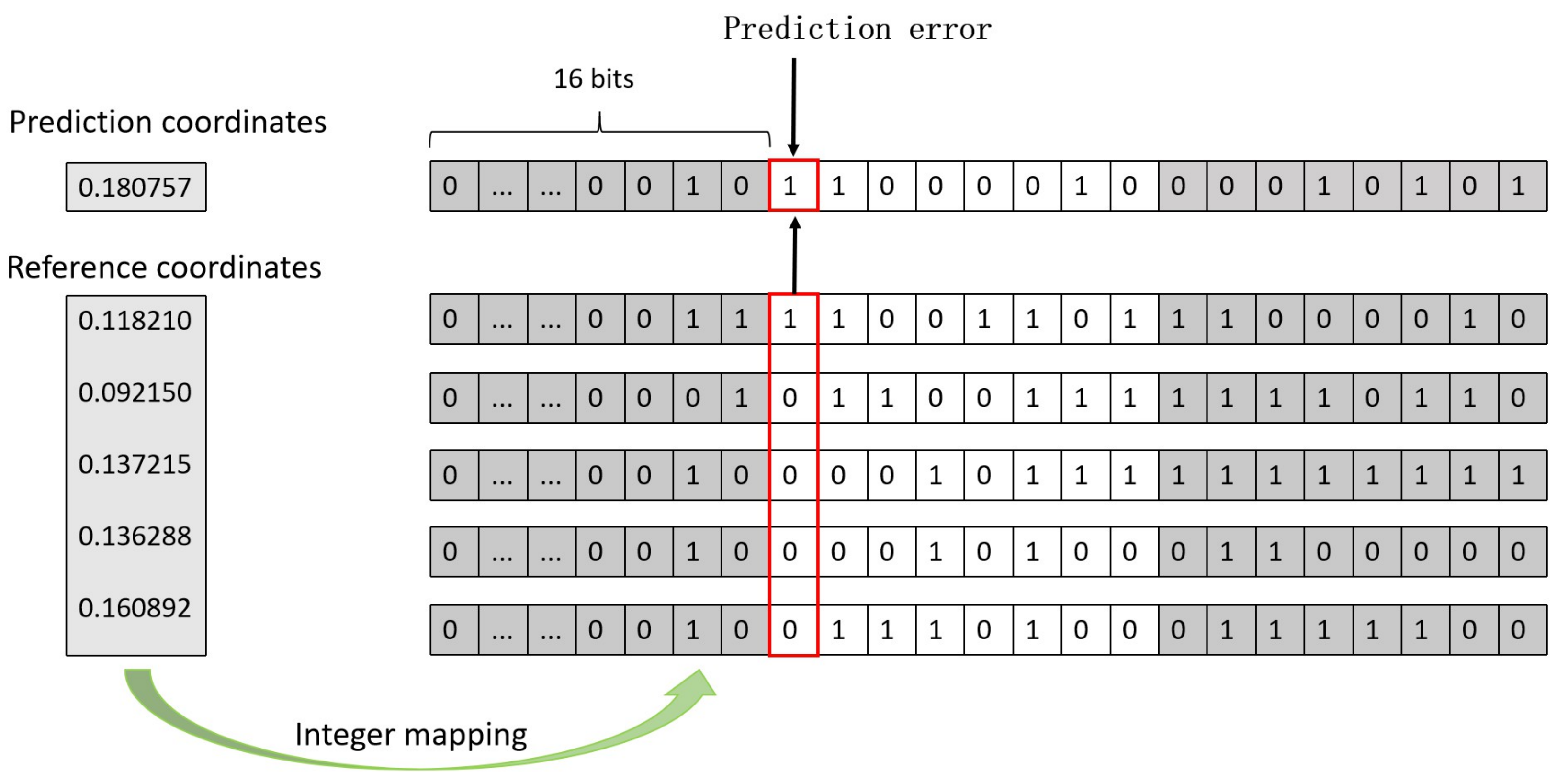}
\end{center}
\caption{An example of prediction error detection test on cow mesh.} \label{fig4}
\end{figure*}
An example shows that when the \emph{${m}$} is 6, the process of selecting the maximum embedding length \emph{${L}$} without prediction error is as follows:
We predict each bit of the prediction coordinates in the order of reference coordinates from MSB to LSB until a certain bit has prediction error.
As shown in Fig. 4, the \emph{${x}$} coordinates of an ``embedded" vertex numbered 1 has the MSB 0. Sender counts the number of 0 and 1 occurrences of the MSB of the ``reference" vertex coordinates numbered 2, 3, 4, 5, 7, 8. if the number of 0s is greater than or equal to the number of 1s, the MSB of the ``embedded" vertex coordinates numbered 1 is predicted to be 0.
Then, the prediction of 2-MSB and 3-MSB were counted until the maximum embedding length \emph{${k1}$} was found. According to Fig. 4, it can be found that the prediction error occurred in the bit prediction when embedding length \emph{${t1}$}=17. Therefore, it can be concluded that the maximum embedding length \emph{${t1}$} was 16 of \emph{${x}$} coordinates when the \emph{${m}$} was 6 on the cow test mesh.
The maximum embedding length of vertex coordinate \emph{${x}$} is calculated as \emph{${t1}$}. Similarly, we calculate the maximum embedding length of vertex coordinate ${y}$, ${z}$ axis \emph{${t2}$} and \emph{${t3}$}. At this time, the maximum embedding length of this vertex is min$\{$\emph{${t1}$},\emph{${t2}$},\emph{${t3}$}$\}$. 
In the data embedding stage, the final maximum embedding length \emph{${L}$} of the embedded vertex is the maximum embedding length of all vertex coordinates.

After the prediction error detection of the \emph{${x}$}-axis, \emph{${y}$}-axis and \emph{${z}$}-axis coordinates of the vertex coordinates, if we get \emph{${n}$} $\ge$ 1, we call vertex numbered 1 as the ``embedded" vertex without prediction error in \emph{C}. Otherwise, the vertex index information will be recorded as auxiliary information. The sender sends auxiliary information together with  marked encrypted mesh to recipient.

\subsection{Encryption}
The original 3D mesh is encrypted by an encryption key \emph{Ke}. After the vertex coordinates are pre-processed, the sender uses Eq. (4) to translate the integer coordinates to binary.
\begin{equation}
\emph{${b}$$_{i,j,u}$}=\lfloor \emph{$\bar{v}$$_{i,j}$}/  2^\emph{${u}$}\rfloor\quad\emph{mod}\quad2,\qquad\emph{${u}$}=0, 1\ldots\emph{${l}$}-1,
\end{equation}
where $\lfloor$$.$$\rfloor$ is a floor function and ${l}$$\le$\emph{${i}$}$\leq$\emph{${N}$} and \emph{${j}$}$\in$$\{$\emph{${x,y,z}$}$\}$, the \emph{${l}$} of the coordinate can be obtained by Eq. (3).

The sender uses a stream cipher function to generate pesudo-random bits \emph{${c}$$_{i,j,u}$}, and encrypts the bitstream of the original 3D mesh \emph{${b}$$_{i,j,u}$} to get the encrypted coordinate binary stream \emph{${e}$$_{i,j,u}$}.
\begin{equation}
\emph{${e}$$_{i,j,u}$}=\emph{${b}$$_{i,j,u}$} \oplus \emph{${c}$$_{i,j,u}$},
\end{equation}
where $\oplus$ stands for exclusive OR.

We can get the encrypted integral mesh using Eq. (6)
\begin{equation}
\emph{${E}$$_{i,j}$}=\sum_{u=0}^{l-1}\emph{${e}$$_{i,j,u}$} \times 10^\emph{${m}$},
\end{equation}
where \emph{${E}$$_{i,j}$} are the integral value of coordinates.
\subsection{Data hiding}
To prevent additional data from being detected, the data hiding key \emph{Kw} is used to encrypt the to-be-embedded data. Sender first calculates ``embedded" set \emph{C} and embed the data into the ``embedded" vertex without prediction errors. Finally, the \emph{n}-MSB of \emph{x}, \emph{y}, and \emph{z} coordinate values are substituted by \emph{${n}$} bit. That is to say, each vertex in ``embedded" set \emph{C} can be embedded with 3$\times$\emph{${n}$} bits using Eq. (7). After data embedding stage, sender gets the encrypted mesh with additional data, i.e., marked encrypted mesh \emph{${E(M)w}$}.
\begin{equation}
\emph{{v$_{i,j}$}$^\emph{$\prime$$\prime$}$}=\emph{s$_{1}$}\times 2^\emph{${l}$-1}+\emph{s$_{2}$}\times 2^\emph{${l}$-2}+\ldots+\emph{s$_{n}$}\times 2^\emph{${l}$-${n}$}+\emph{{v$_{i,j}$}$^\emph{$\prime$}$}mod 2^{\emph{l}-(n+1)},
\end{equation}
Where \emph{s$_{k}$} is additional data and \textrm{${l}$$\le$\emph{${k}$} $\leq$ ${n}$}, \emph{${n}$} is embedding length and \textrm{1$\le$\emph{${n}$} $\leq$ ${L}$}, {\emph{{v$_{i,j}$}$^\emph{$\prime$}$}}$\in$\emph{C} is the vertex after preprocessing and encryption, \emph{{v$_{i,j}$}$^\emph{$\prime$$\prime$}$} is the vertex of marked encrypted mesh.
\begin{figure*}[!ht]
  \centering
  \subfigure[]{
   \label{fig-5-a}
    \includegraphics[width=0.23\textwidth]{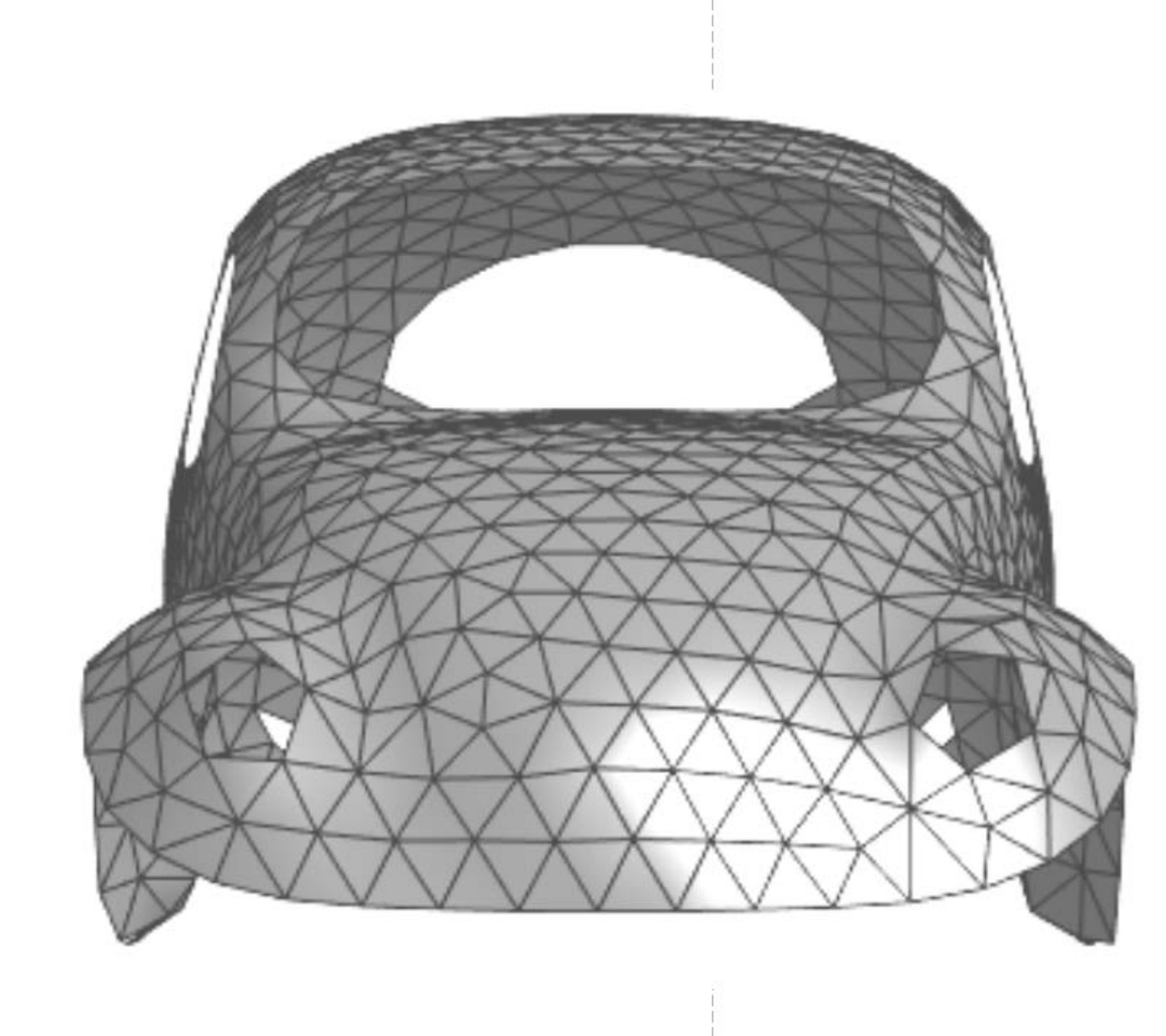}
  }
   \subfigure[]{
   \label{fig-5-b}
    \includegraphics[width=0.26\textwidth]{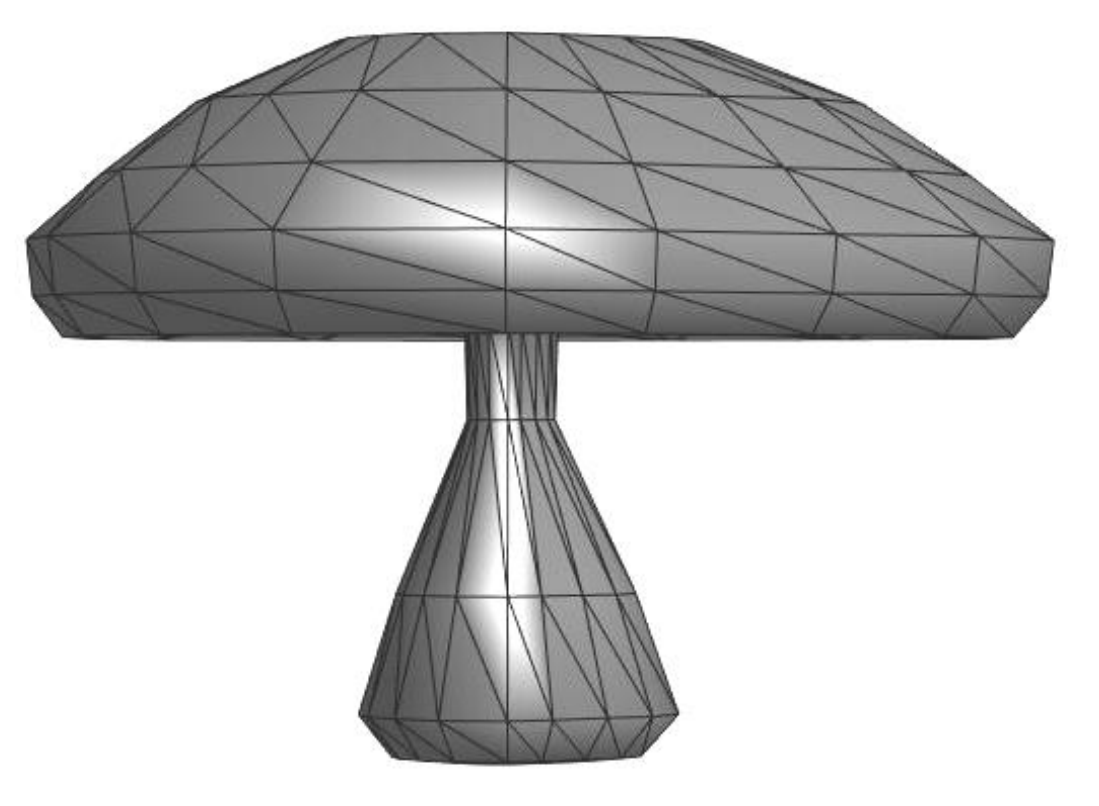}
  }
   \subfigure[]{
   \label{fig-5-c}
    \includegraphics[width=0.15\textwidth]{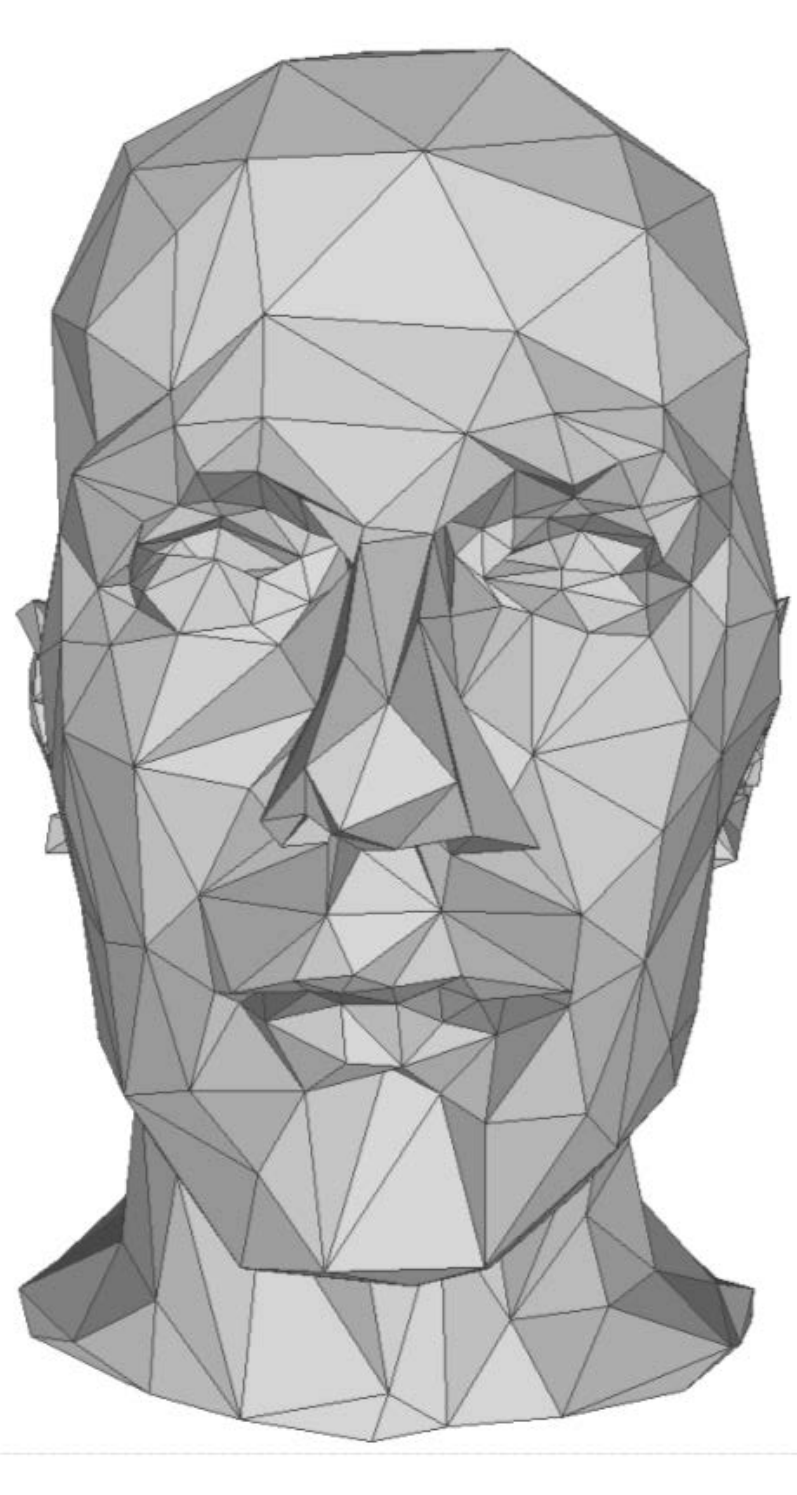}
  }
   \subfigure[]{
   \label{fig-5-d}
    \includegraphics[width=0.20\textwidth]{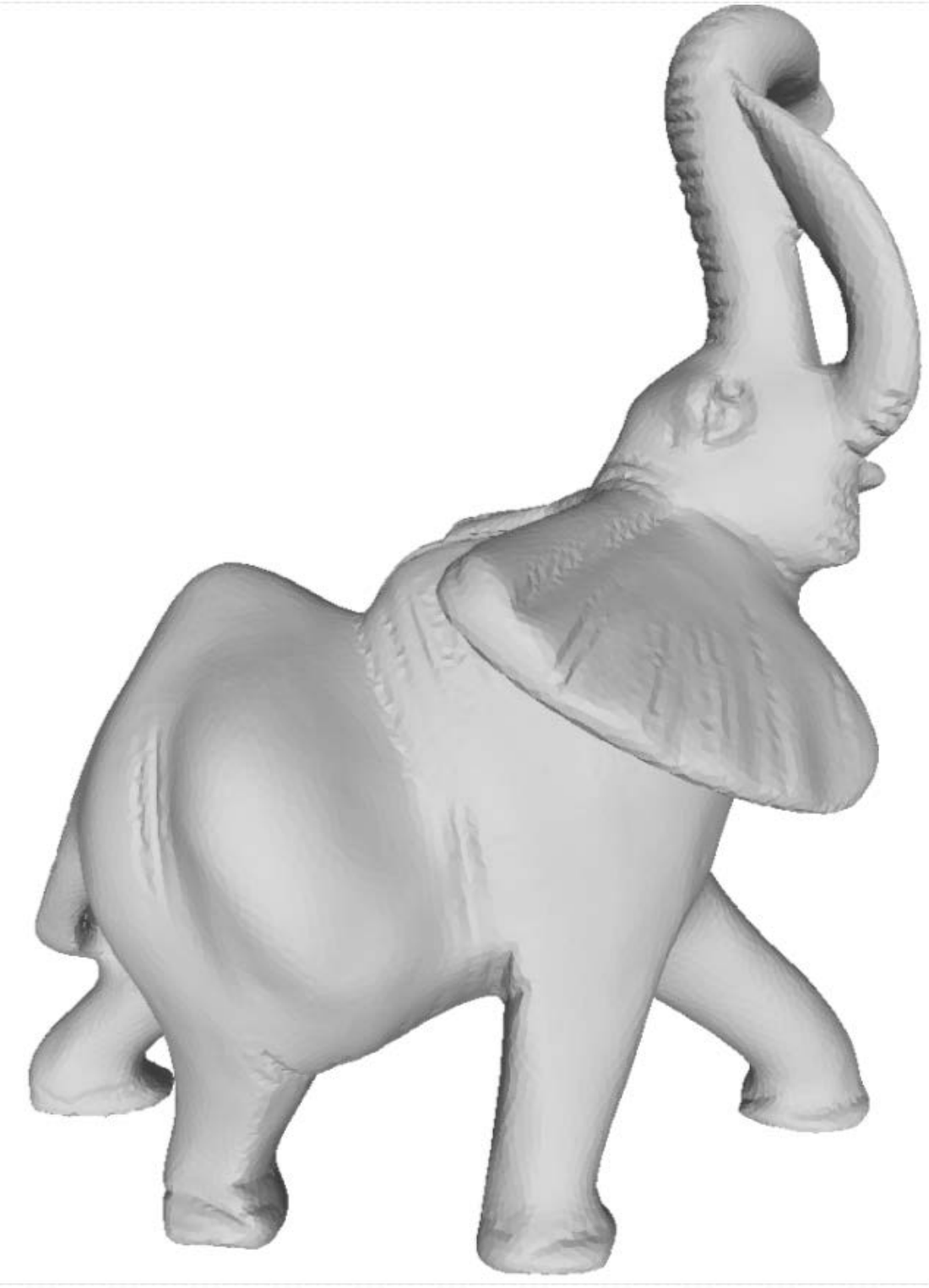}
  }
  \caption{ Test Meshes: (a) Beetle, (b) Muchroom, (c) Mannequin, (d)  Elephant.}
\label{fig5}
\end{figure*}
\subsection{Data Extraction and Mesh Recovery} 
After receiving the marked encrypted mesh ${E(M)w}$, since our method is separable, the recipient can use the data hiding key ${Kw}$ to extract the additional data and use the encryption key ${Ke}$ to recover the original mesh separately. There are three cases which are as follows: 
\subsubsection{Extraction with only Data Hiding Key} 
In case 1, the $n$-MSB is extracted from the vertex coordinates of ``embedded" set \emph{C} without prediction errors, and then the corresponding plaintext additional data is obtained by using the data hiding key \emph{Kw}.
\begin{equation}
\emph{s$_{k}$} =\emph{{v$_{i,j}$}$^\emph{$\prime$$\prime$}$}/2^\emph{$l$-$k$}, \qquad  \textrm{$1$$\le$\emph{$k$} $\leq$ $n$}
\end{equation}
where \emph{{v$_{i,j}$}$^\emph{$\prime$$\prime$}$}$\in$ \emph{C} is vertex of the marked encrypted mesh.
\subsubsection{Mesh Recovery with only Encryption Key} 
In case 2, the recipient can recover the marked encrypted mesh $E(M)w$ to get the original mesh $M$. The original mesh $M$ is recovered in two steps : mesh decryption and Multi-MSB prediction recovery.

The pseudorandom bits \emph{${c}$$_{i,j,u}$} are generated by the encryption key $Ke$, and used to perform xor function with \emph{${e}$$_{i,j,u}$$^\emph{$\prime$$\prime$}$} to decrypt the marked encrypted mesh $E(M)w$.

\begin{equation}
\emph{${b}$$_{i,j,u}$$^\emph{$\prime$$\prime$}$}= \emph{${e}$$_{i,j,u}$$^\emph{$\prime$$\prime$}$}\oplus \emph{${c}$$_{i,j,u}$},
\end{equation}
Where \emph{${e}$$_{i,j,u}$$^\emph{$\prime$$\prime$}$} is the binary stream of the marked encrypted mesh, \emph{${b}$$_{i,j,u}$$^\emph{$\prime$$\prime$}$} is the binary stream of the decrypted mesh with additional data and \emph{$u$}=0, 1\ldots\emph{$l$}-1.

After decrypting the mesh, the vertex coordinates of \emph{R} have been recovered. 
However, in the data embedding stage, the $n$-MSB of the coordinates of the set of vertices in \emph{C} is replaced by the additional data. After the mesh is decrypted, due to the spatial correlation of the original mesh, recipient can predict the $n$-MSB of the ``embedded" vertex by the $n$-MSB of the adjacent vertices around the ``embedded" vertex. The use of adjacent reference coordinates to predict embedded vertex coordinates is called ring- prediction. That is to say, recipient can obtain high quality recovered mesh by using ring-prediction in the recovery stage.
As for Fig. 4, the coordinate values of adjacent vertices 2, 3, 4, 5, 7 and 8 have been correctly restored after decryption. Based on their $n$-MSB values, the $n$-MSB of vertex numbered 1 are predicted to be 0 or 1. For example, when predicting the MSB of v$_{1,\emph{x}}$, we count the MSB of \emph{$x$} coordinates of vertex index numbers 2, 3, 4, 5, 6. If the highest bit 0 occurs more than or equal to the number of times 1 occurs, then the MSB of v$_{1,\emph{x}}$ is predicted to be 0. Using the same prediction method, recipient can correctly recover n-MSB of embeddable vertices.
\subsubsection{Extraction and Mesh Recovery with Both Keys} 
In case 3, if the recipient has both the data hiding key $Kw$ and the encryption key $Ke$, the recipient can extract the additional data and recover the original 3D mesh perfectly. Note that data extraction step needs to be performed before mesh restoration. 

\begin{table*}[!ht]
\newcommand{\tabincell}[2]{\begin{tabular}{@{}#1@{}}#2\end{tabular}}
\vspace{-0.2cm}
 \renewcommand\arraystretch{1.3}
\caption{The embedding capacity of the tested meshes on different $m$ and embedding length $n$.}
\centering 
\setlength{\tabcolsep}{0.4mm}
\begin{tabular}{@{}lccccccc ccccccc ccccccc c @{}}
 \toprule
Meshes & m/n &1&2 &3& 4&5 &6        &7 &8 &9 &10 &11 &12 &13 &14 &15 &16 &17 &18 &19 &\ldots &32 \\ \midrule
\multirow{8}{*}{Beetle}                 & 2          &0.98     &1.97              & 3.03   & 3.57& 3.70                    &1.78 
&0.06 &0.02   &-  &- &- &-  &-  &-  &-  &- &- &-  &-  &\ldots &-
\\ \cmidrule(l){2-23} 
&  3         & 0.98     &1.97            &2.96        & 3.94         &4.93  &5.92    
&6.90 &7.74 &7.59 &6.64 &1.13 &0.10  &0.03  &0.04 &0  &0  &-  &-  &-  &\ldots &- 

\\ \cmidrule(l){2-23} 
 & 4 &0.98     &1.97             &2.96       &3.87         &4.70 &5.08    
&4.01 &0.19 &0.05 &0 &0 &0  &0  &0  &0  &0  &0  &- &-  &\ldots &- 
\\ \cmidrule(l){2-23} 
&5& 0.98    &1.97           &2.96      & 3.95      &4.94  &5.93    
&6.92 &7.91&8.90 &9.89 &10.88 &11.87  &12.86  &13.85  &14.84  &16.51  &0  &0  &0  &\ldots &0
\\ \cmidrule(l){2-23} 
&6& 0.98    &1.97           &2.96      & 3.95      &4.94  &5.93   
&6.92 &7.91&8.90 &9.89 &10.88 &11.87  &12.86   &13.64  &13.11  &10.97  &1.65  &0.38  &0  &\ldots &0
\\ \cmidrule(l){2-23} 
&7& 0.98    &1.97           &2.96      & 3.95      &4.94  &5.93   
&6.92 &7.91&8.90 &9.68 &10.28&9.98  &7.30  &0.34  &0.04  &0  &0  &0  &0  &\ldots &0
\\ \cmidrule(l){2-23} 
&8& 0.98    &1.97           &2.96      & 3.95      &4.94  &5.93      
&6.80 &7.31 &6.72 &3.76 &0.10 &0.03  &0 &0  &0 &0  &0  &0  &0  &\ldots &0 
\\ \cmidrule(l){2-23} 
&9& 0.98    &1.97           &2.96        & 3.87        &4.31  &4.29    
&1.06 &0.12 &0 &0 &0 &0  &0  &0  &0  &0  &0 &0  &0  &\ldots &0 
                 \\  \toprule
\multirow{8}{*}{Muchroom}                   &  2     &1.11    &1.91                     &1.79 &1.27 &0.53  &0.07                
&0 &0&- &- &-  &- &- &-   &- &- &-  &- &-  &\ldots &-  \\ \cmidrule(l){2-23} 
                                          &   3     & 1.11    &2.23               &3.34       &4.46      & 5.57  &6.69  
&5.76 &3.39 &0.71 &0.13 &0 &0  &0  &0  &0  &0  &- &- &-  &\ldots &- \\ \cmidrule(l){2-23} 
&4   &1.11    &2.23                 & 3.14                             & 3.23    & 1.65       &0.23    
&0.09&0  &0  &0 &0 &0  &0  &0  &0  &0  &- &- &-  &\ldots &-                 
\\ \cmidrule(l){2-23} 
&5& 1.11    &2.23           &3.34      & 4.46     &5.57  &6.69    
&7.80 &8.92&10.03 &11.15 &12.26&13.38  &14.49  &15.61  &16.72  &15.71  &10.38  &5.25  &1.51  &\ldots &0
\\ \cmidrule(l){2-23} 
&6& 1.11    &2.23           &3.34      & 4.46     &5.57  &6.69    
&7.80 &8.92&10.03 &11.15 &12.26&13.38   &10.87  &6.13  &1.39  &0.21  &0&0 &0  &\ldots &0
\\ \cmidrule(l){2-23} 
&7& 1.11    &2.23           &3.34      & 4.46     &5.57  &6.69  
&7.80 &8.92&9.43 &7.96&3.65  &0.47  &0.17  &0  &0 &0  &0  &0  &0  &\ldots &0

\\ \cmidrule(l){2-23} 
&8& 1.11    &2.23           &3.34      & 4.46     &5.57  &5.57    
&4.27 &2.54&0.83 &0 &0&0  &0  &0  &0 &0  &0  &0  &0  &\ldots &0

\\ \cmidrule(l){2-23} 
&9& 1.11    &2.23           &2.58        & 1.80       &0.79  &0.23    
&0 &0 &0 &0 &0 &0  &0  &0  &0  &0  &0 &0  &0  &\ldots &0 \\ \toprule

\multirow{8}{*}{Mannequin}                 &  2     &  0.94 & 1.64 &  1.42&  0.89&  0.21&  0.04&  0&  0&- &- &-  &- &- &- &- &- &- &- &- &  \ldots&-

 \\ \cmidrule(l){2-23} 
                                          &   3          &  0.94 & 1.89 &  2.83&  3.78&  4.73&  5.38&  5.10&  3.19&  1.07&  0.21 & 0 & 0 &  0&  0&  0&  0&  -&  -&-&  \ldots&  -   \\ \cmidrule(l){2-23} 
                                          & 4      & 0.94 & 1.83 &  2.37&  2.07&  1.05&  0.12&  0&  0&  0&  0 & 0 & 0& 0&  0& 0&  0&  -&  -&-&  \ldots& -                  
\\ \cmidrule(l){2-23} 
&5 &  0.94 & 1.89 &  2.83&  3.78&  4.73&  5.67&  6.62&  7.57&  8.51&  9.46& 10.40  & 11.35 &  12.30&  13.24&  13.66&  12.7&  5.38&  0.66&0&\ldots  &  0
\\ \cmidrule(l){2-23} 
&6 &  0.94 & 1.89 &  2.83&  3.78&  4.73&  5.67&  6.62&  7.57&  8.51&  9.46 & 10.40 & 10.85 &  9.65&  6.08&  1.89&  0.11&  0&  0&0&  \ldots&  0
\\ \cmidrule(l){2-23} 
&7 & 0.94 & 1.89&  2.83&  3.78&  4.73&  5.67&  6.62&  7.28&  6.93&  5.18 & 2.08 & 0.50 &  0.09&  0&  0&  0&  0&  0&0&  \ldots&  0
\\ \cmidrule(l){2-23} 
&8 &  0.94  & 1.89 &  2.83&  3.78&  4.59&  4.92&  3.18&  1.51&  0.31&  0 & 0 & 0 &  0&  0&  0&  0&  0&  0&0&  \ldots& 0
\\ \cmidrule(l){2-23} 
&9 &  0.94  & 1.79 &  2.12&  1.51&  0.49&  0.04&  0&  0&  0&  0 & 0 & 0 &  0&  0&  0&  0&  0& 0&0&  \ldots&0

 \\ \toprule
\multirow{8}{*}{Elephant} 

&2          & 1.02       &2.04               &3.05                    &3.98                        & 4.76  &5.24
&5.07 &2.44   &- &- &- &- &- &- &- &- &-  &- &-  & \ldots &- \\ \cmidrule(l){2-23}  
\multicolumn{1}{c}{}                      

&3        & 1.02       &2.04                &3.06       &4.08          & 5.10  &6.12  
&7.14 &8.07  &8.82  &9.32 &9.16 &7.93  &1.47  &0.09 &0.001  &0  &- &- &-  &\ldots &-   \\ \cmidrule(l){2-23} 
\multicolumn{1}{c}{}                      

& 4       &1.02    &2.04   &3.06  & 4.07    &5.01       &5.80       &6.32 &6.28   &4.81  &0.28 &0.02 &0  &0  &0  &0 &0  &0  &0  &0  &\ldots &- 
\\ \cmidrule(l){2-23}

&5  &1.02    &2.04   &3.06      & 4.08      &5.10 &6.13    
&7.15  &8.17  &9.19   &10.21 &11.24 &12.26  &13.28  &14.30  &15.32  &16.34  &17.29  &17.92  &18.12  &\ldots &0
\\ \cmidrule(l){2-23} 

&6&1.02    &2.04   &3.06      & 4.08      &5.10  &6.13    
&7.15  &8.17  &9.19   &10.21 &11.24 &12.26  &13.28  &14.30  &15.32  &16.34  &17.29  &17.92  &18.12  &\ldots &0

\\ \cmidrule(l){2-23} 
&7&1.02    &2.04   &3.06      & 4.08      &5.10  &6.13    
&7.15  &8.17  &9.19   &10.21 &11.24 &12.26  &13.28   &14.14  &14.71  &14.92  &14.17  &11.88  &2.40  &\ldots &0

\\ \cmidrule(l){2-23} 
&8&1.02    &2.04   &3.06      & 4.08      &5.10  &6.13   
&7.11 &7.98 &8.62 &8.81 &8.20 &4.35  &0.15  &0.02  &0 &0  &0  &0  &0  &\ldots &0

\\ \cmidrule(l){2-23} 
&9&1.02    &2.04   &3.06      & 4.05         &4.92  &5.62    
&5.95 &5.52 &1.35 &0.07 &0 &0  &0  &0  &0  &0  &0  &0  &0  &\ldots &0

\\ \toprule
\end{tabular}
\end{table*}
\section{EXPERIMENTAL RESULTS AND DISCUSSION}
In this section, the reversibility and embedding capacity of the improved method are analyzed, and the results are compared with the state-of-the-art methods~\cite{ref_article17}~\cite{ref_article18}.
As shown in Fig. 5, four standard meshes: Beetle, Mushroom, Mannequin, Elephant. Two datasets: 1815 meshes with off format from The Princeton Shape Retrieval and Analysis Group \footnote{http://shape.cs.princeton.edu/benchmark/index.cgi.} and those in OBJ format from The Stanford 3D Scanning Repository\footnote{http://graphics.stanford.edu/data/3Dscanrep/.} were used to test performance.
The key indicator is embedding rate (ER) presented by bpv (bits per vertex). We analyze the embedding capacity of proposed method and compare with the state-of-the-art methods in section A.
The data embedding process causes distortion to the original mesh models, which cannot be observed by the naked eye. 
Hausdorff distance and signal-to-noise ratio (SNR) are used to evaluate the reversibility of the method in section B.
The additional data embedded is a randomly generated 0/1 sequence.
\vspace{-0.2cm}  
\subsection{Embedding Capacity}
The embedding rate is measured by the number of bits per vertex (bpv), which is the ratio of the number of embedded bits to the number of vertices in the mesh model. 
This section mainly analyzes the influence of \emph{m} and embedding length \emph{n} on the embedding capacity. 
For “Beetle”, Table II showes that when \emph{m} =5 and the \emph{n}=16, the maximum embedding rate is 16.51bpv.
Similarly, when m=5 and n=15, the maximum embedding rate of Mushroom is 16.72 bpv; when m=5 and n=15, the maximum embedding rate of Mannequin  is 13.66 bpv; when m=5 and n=18, the maximum embedding rate of Elephant is 18.12 bpv.
Therefore, the maximum embedding rate of Beetle is 16.51 bpv, Muchroom is 16.72 bpv, Mannequin is 13.66 bpv and Elephant is 18.12 bpv.
\begin{figure*}[!ht]
  \centering
\vspace{-0.3cm}  
\setlength{\belowcaptionskip}{-0.2cm}  
     \subfigure[]{
   \label{fig-6-a}
    \includegraphics[width=0.45\textwidth]{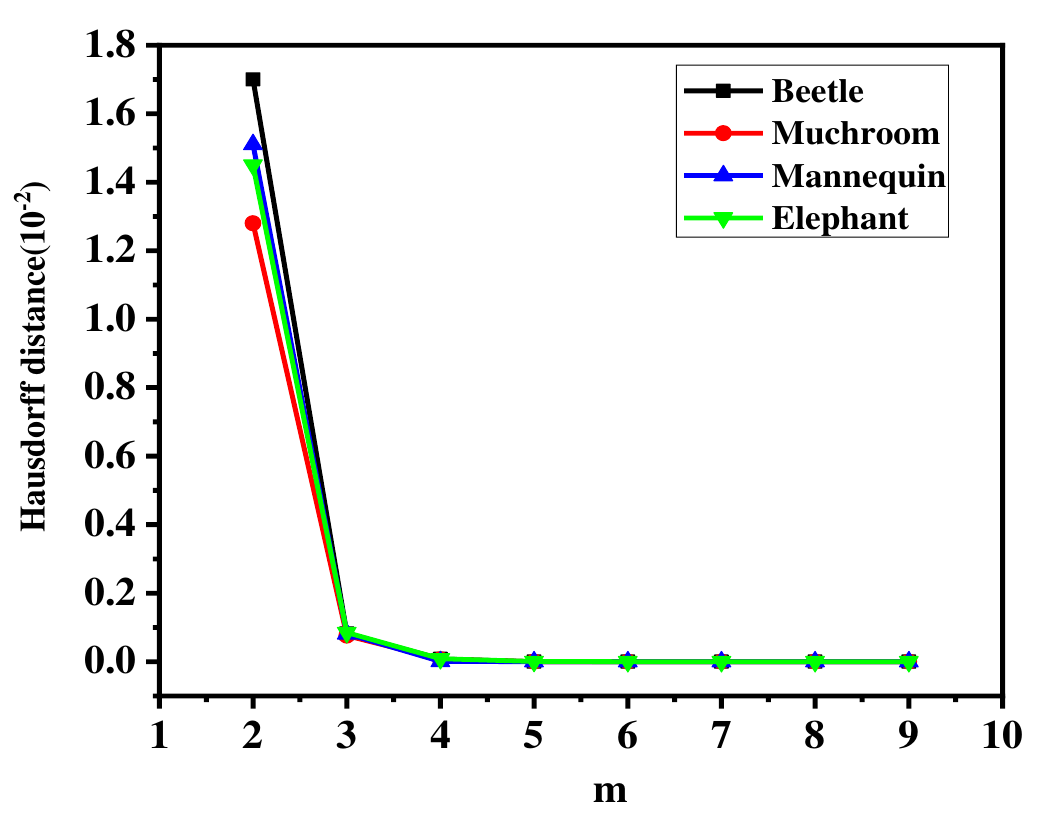}
  }
   \subfigure[]{
   \label{fig-6-b}
    \includegraphics[width=0.45\textwidth]{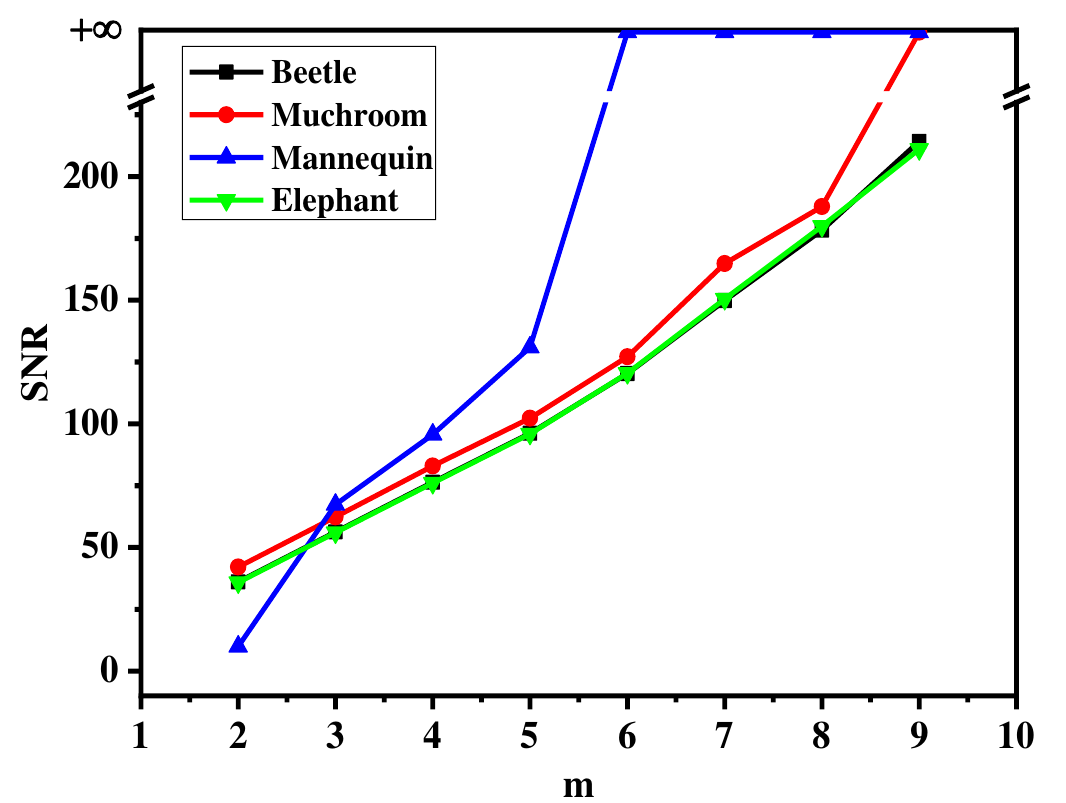}
  }
  \caption{ Results of four test meshes on different accuracy \emph{m} under maximum embedding capacity.
(a) Hausdorff distance, (b) SNR.}
\label{fig6}
\end{figure*}
\subsection{Geometric and Visual Quality} 
Hausdorff distance and signal-to-noise ratio (SNR) can be used to measure the geometric distortion of the meshes.
Hausdorff distance is a measure describing the similarity between two sets of points, which is a definition of the distance between two sets of points. Assuming there are two sets of \emph{A}=(\emph{a$_1$},\emph{a$_2$}\ldots\emph{a$_p$}) and \emph{B}=(\emph{b$_1$},\emph{b$_2$}\ldots\emph{b$_q$}), the Hausdorff distance between these two sets of points is defined as:
\begin{equation}
H(A,B)=max(\emph{h}(A,B),\emph{h}(B,A)),
\end{equation}
\begin{equation}
\emph{h}(A,B)=max(\emph{a}\in A)min(\emph{b}\in B) \parallel\emph{a}-\emph{b}\parallel,
\end{equation}
\begin{equation}
\emph{h}(B,A)=max(\emph{b}\in B)min(\emph{a}\in A) \parallel\emph{b}-\emph{a}\parallel,
\end{equation}
Where $\parallel$$.$$\parallel$ is the distance between point a of set A and point b of set B (such as L2), \emph{p} and \emph{q} are the number of elements in the set.

The geometrical distortion of a mesh after adding some noise to the mesh content can be measured by the the signal-to-noise ratio (SNR), which is defined as follows:
\emph{SNR}=
\begin{equation}
10 \times \lg \frac{ \sum_{i=1}^{N} [(v_{i,x}-\overline v_x)^2+(v_{i,y}-\overline v_y)^2+(v_{i,z}-\overline v_z)^2]}{ \sum_{i=1}^{N} [(g_{i,x}-\overline v_x)^2+(g_{i,y}-\overline v_y)^2 +(g_{i,z}-\overline v_z)^2 ] },
\end{equation}

Where  $\bar{v}$$_x$,$\bar{v}$$_y$,$\bar{v}$$_z$ are the average of the mesh coordinates, v$_{i,x}$ ,v$_{i,y}$ ,v$_{i,z}$ are the original coordinates, g$_{i,x}$, g$_{i,y}$, g$_{i,z}$ are the modified mesh coordinates
 value, $N$ is the number of vertices.

The value of $m$ is a trade-off between the quality of the recovered mesh and the computational overhead of the process. In the case of maximum embedding capacity, we tested Hausdorff distance and SNR between the original mesh and  recovered mesh with change of $m$. When m is 1, the decimal retention accuracy is too low, resulting in poor quality of the recovered mesh, which is not suitable for most application scenarios and has no practical significance. Therefore, we varied $m$ from 2 to 9 to calculated the Hausdorff distance and SNR between  original and recovered mesh and observe the phenomenon. 

As shown in Fig. 6, the Hausdorff distance decreases linearly with the increase of \emph{m}, while the SNR increases linearly, indicating that the quality of the recovered mesh becomes higher with the increase of \emph{m}. 
We can observe that m=4 balances the quality of the recovered meshes and the overhead time.
Therefore, we choose m=4 for experiments and Fig. 8. shows experimental results, demonstrating the visual effects of the original mesh at different stages of the proposed method. However, the naked eye cannot notice the difference between the original and recovered mesh due to the reason that the recipient with encryption key $Ke$ can prefectly recover the original mesh. That is, the proposed method doesnot introduce perceptual distortion. 
\subsection{Performance Comparison} 
The performance of the proposed framework is compared with the state-of-the-art~\cite{ref_article17} and~\cite{ref_article18}, this part gives the performance compasision in terms of bpv, Hausdorff distance and SNR . As shown in Fig. 7(a), the maximum embedding rate obtained by the improved method is significantly higher than that obtained methods in~\cite{ref_article17} and~\cite{ref_article18}.
Moreover, in order to reduce the influence caused by the random selection of the test meshes, we test the performance of embedding capacity on the The Princeton Shape Retrieval and Analysis Group datasets.
Fig. 7(b) shows that the proposed method has higher embedding capacity compared to the existing RDH-ED methods.
\begin{figure*}[!ht]
\vspace{-0.2cm} 
  \centering

  \subfigure[]{
   \label{fig-7-a}
    \includegraphics[width=0.45\textwidth]{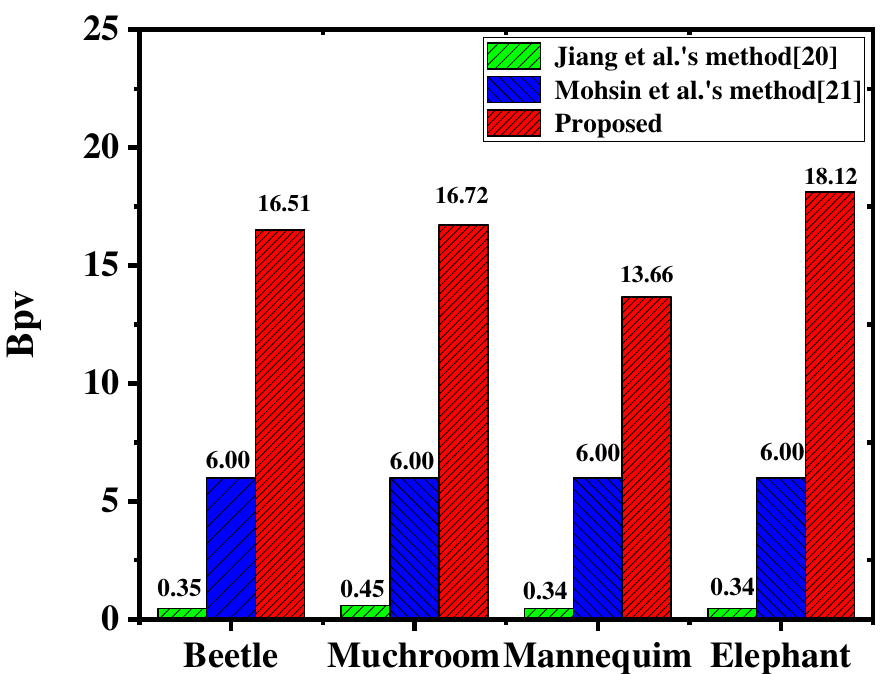}
  }
   \subfigure[]{
   \label{fig-7-b}
    \includegraphics[width=0.43\textwidth]{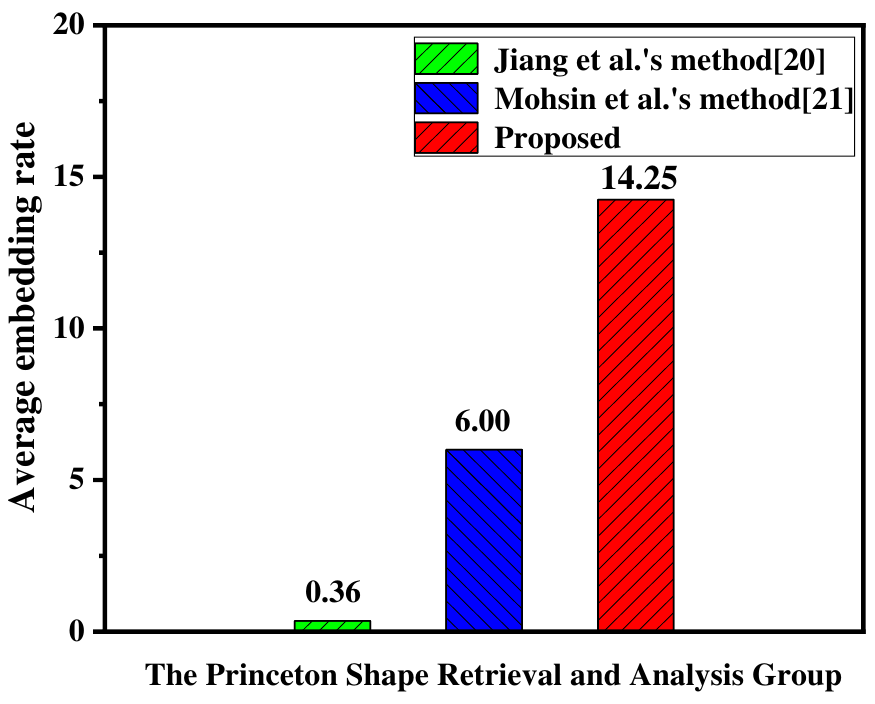}
  }
  
  \caption{ Test Meshes: (a) Comparison of maximal enbedding rates of test meshes between our method and state-of-the-art methods, (b)Average embedding capacity on a data set.}  
\label{fig7}
\end{figure*}
\begin{table*}[]
\newcommand{\tabincell}[2]{\begin{tabular}{@{}#1@{}}#2\end{tabular}}

\caption{\label{tb::tab2} Proposed method compared with the state-of-the-art methods on the four test meshes.}
\centering 
 \renewcommand\arraystretch{1.1}
\setlength{\tabcolsep}{3mm}
\begin{tabular}{@{}lccccccc@{}}
\toprule

Test meshes                               & \multicolumn{1}{l}{Methods} &\tabincell{l}{Embedding \\Rate (bpv)}&\tabincell{l}
{Embedding\\ Capacity (bit)}   &\tabincell{l}{Hausdorff \\distance \emph{${10}$$^\emph{$\emph{-3}$}$}}& \multicolumn{1}{l}{SNR } &\tabincell{l}{Data error\\percent} &Separable\\ \midrule
\multirow{3}{*}{Beetle}                 & [20]            & 0.35  &348              & 9.9000                        & 43.00 & 36.78$\%$                     &NO \\ \cmidrule(l){2-8} 
                                          &  [21]         & 6.00     &5928             &0.0820        & 90.8471         &0  &NO   \\ \cmidrule(l){2-8} 
                                          & \textbf{Proposed}         & 16.51  &16312                & 0.0086                             & 96.20 & 0    &YES

                 \\ \midrule
\multirow{3}{*}{Muchroom}                   &  [20]     & 0.45    &106                     & 10.1000                     & 47.91 &32.08$\%$    &NO                 \\ \cmidrule(l){2-8} 
                                          &   [21]      &  6.00    &1356                & 0.0640        &100.0923      & 0  &NO   \\ \cmidrule(l){2-8} 
                                          & \textbf{Proposed}    & 13.66    &3087                  & 0.0081                             & 102.25                              & 0        &YES                  \\ \midrule

\multirow{3}{*}{Mannequin}                 &  [20]     & 0.34  &146                    & 9.3000                        & 52.47                              & 45.21$\%$   &NO                \\ \cmidrule(l){2-8} 
                                          &   [21]          &  6.00   &2568             &0.0728        & 100.0923         & 0     &NO     \\ \cmidrule(l){2-8} 
                                          & \textbf{Proposed}       & 16.20     &6934                & 0.0040                             & 130.92                              & 0        &YES                         \\ \midrule

\multirow{3}{*}{Elephant} &  [20]          & 0.34       &8582               & 0.1100                     & 41.50                        & 4.94$\%$                     &NO \\ \cmidrule(l){2-8}  
\multicolumn{1}{c}{}                      &    [21]        &  6.00     &149730                & 0.0848        & 90.8037          & 0    &NO     \\ \cmidrule(l){2-8} 
\multicolumn{1}{c}{}                      & \textbf{Proposed}       & 18.12     &452185              & 0.0086                             & 95.97                              &0       &YES      
\\ \toprule
\end{tabular}
\end{table*}
We compare geometric quality of the proposed method with~\cite{ref_article17}and~\cite{ref_article18}. 
Table III showed that method~\cite{ref_article18} and~\cite{ref_article17} are an inseparable scheme, i.e. mesh models decryption and data extraction are carried out at the same time. 
In the data extraction stage, the Jiang et al.'s ~\cite{ref_article17} first decrypts the marked encrypted mesh, and then uses the spatial correlation to predict the least significant bits (LSB) of embedded vertex coordinates to recovery the original mesh. In order to improve the shortcoming of the error rate in data extraction is large of ~\cite{ref_article17}, because MSB prediction is more accurate than LSB prediction during the decoding phase, our method can directly extract additional data from the Multi-MSBs of coordinates of encrypted vertices without any error in ``embedded" set. 
Table III showes that the proposed separable method compared with ~\cite{ref_article17}~\cite{ref_article18} not only improve the embedding capacity, but also obtain a higher quality recovery mesh and achieve error-free extraction of data.

\begin{figure*}[!ht]
\vspace{-0.4cm}
  \centering
    \includegraphics[width=0.15\textwidth]{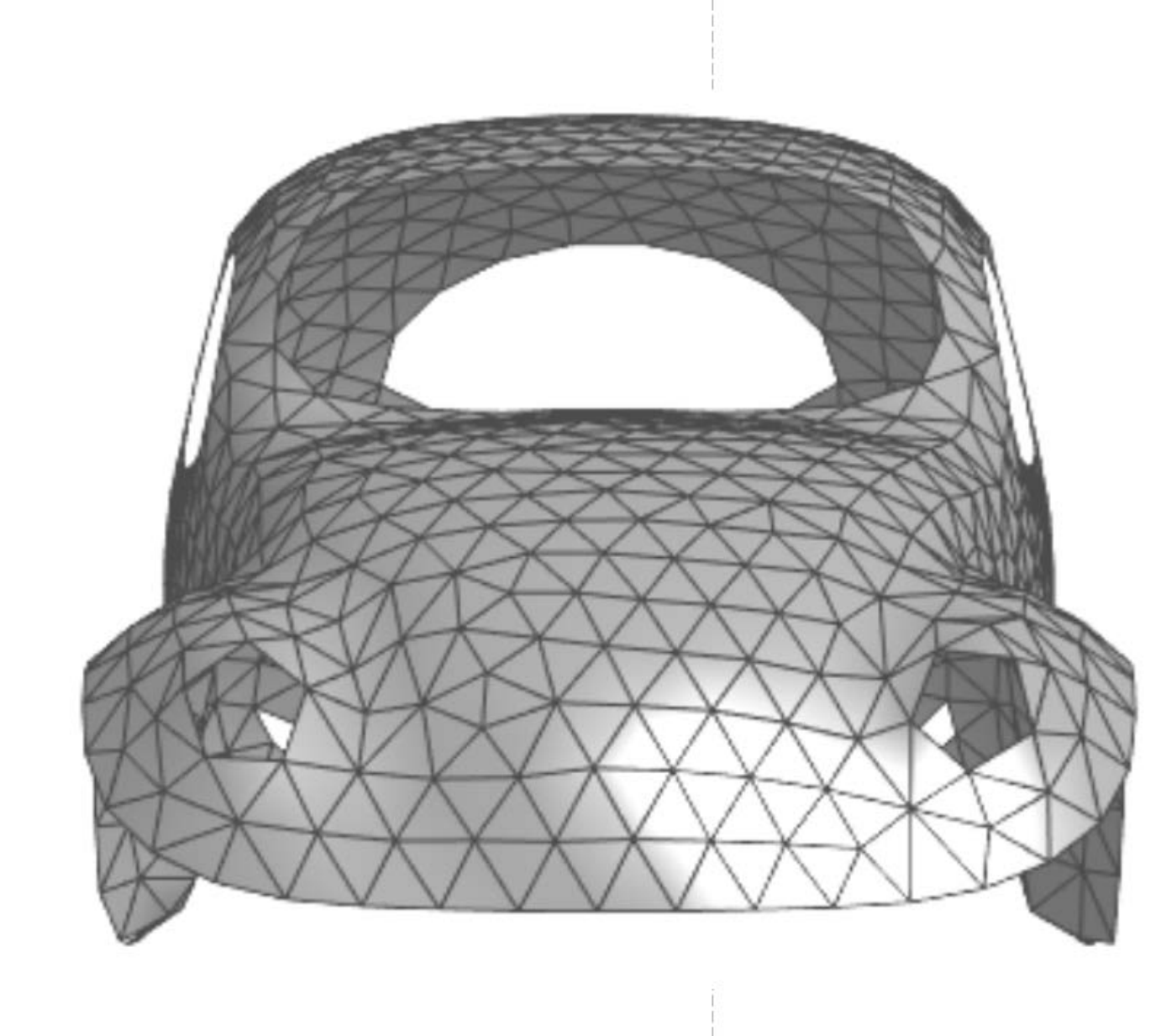}
    \includegraphics[width=0.15\textwidth]{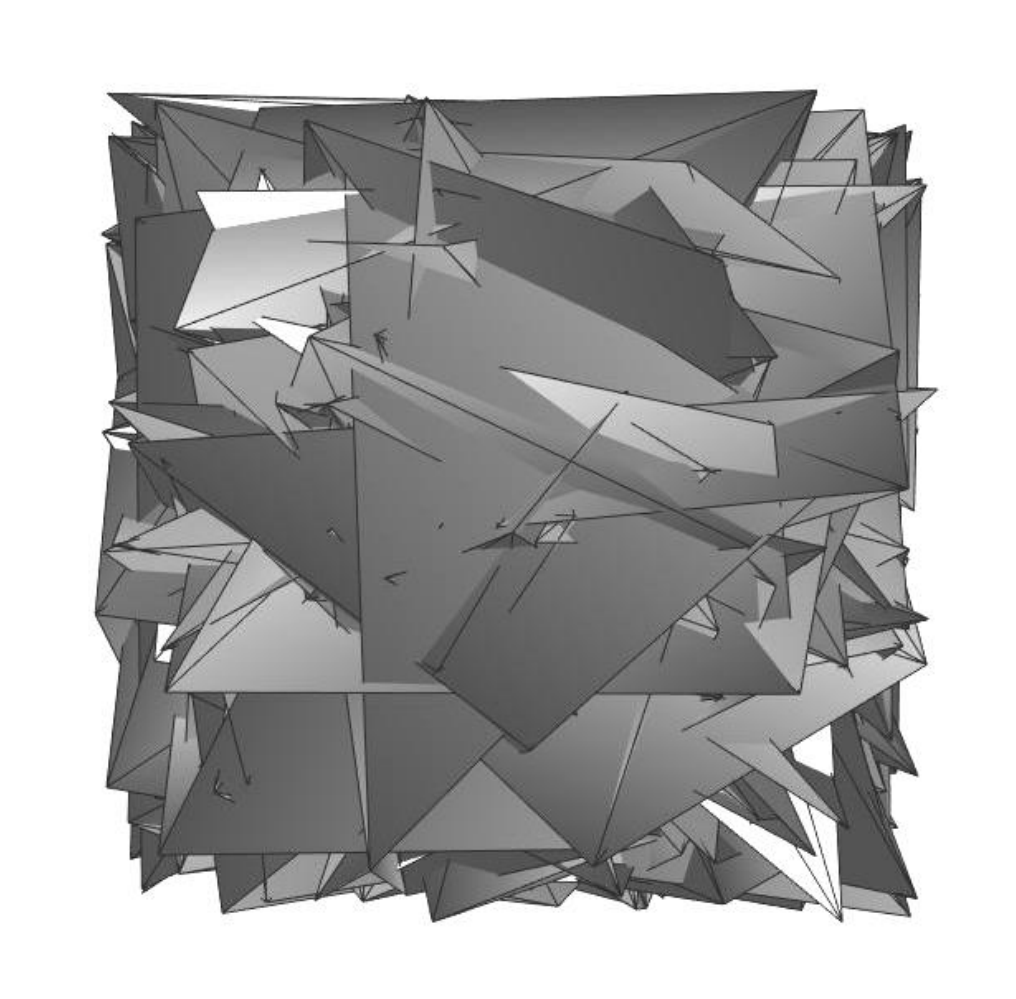}
    \includegraphics[width=0.15\textwidth]{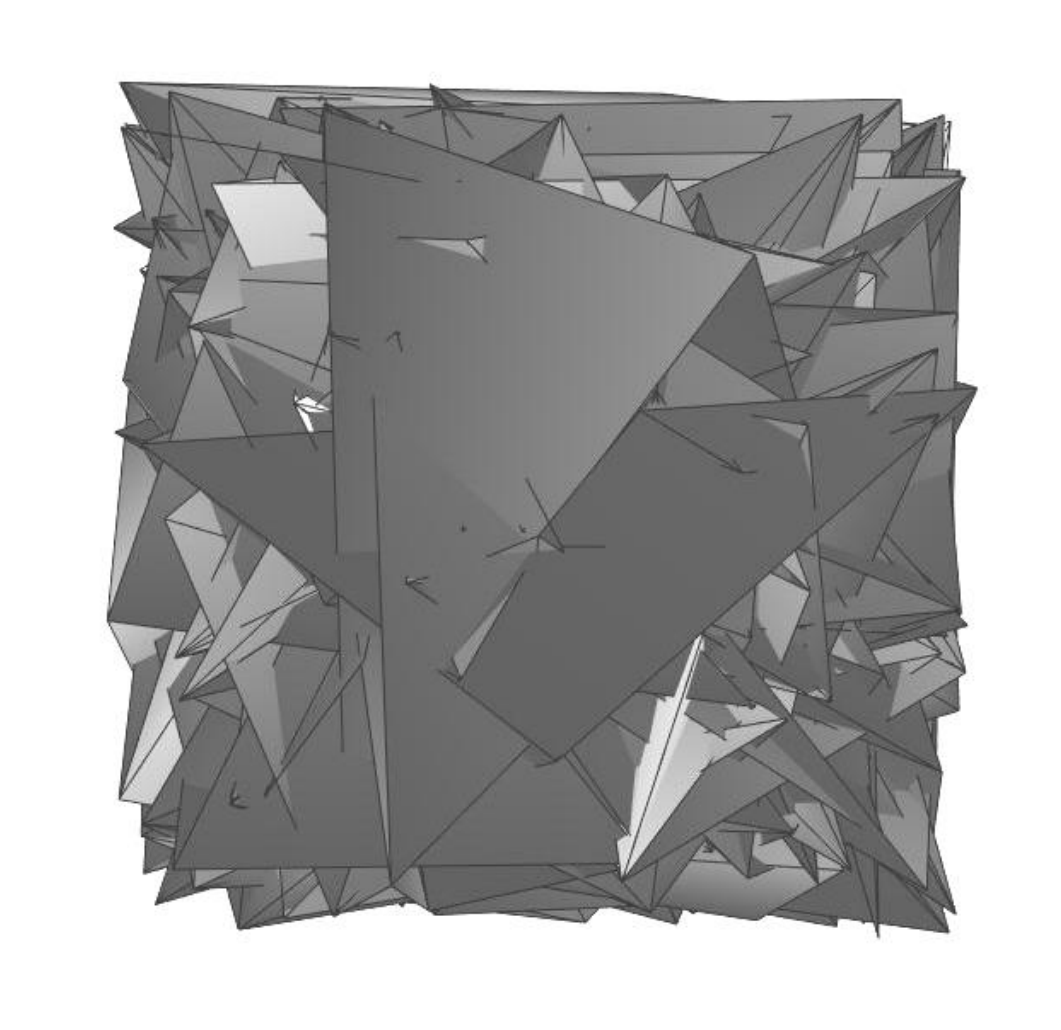}
    \includegraphics[width=0.15\textwidth]{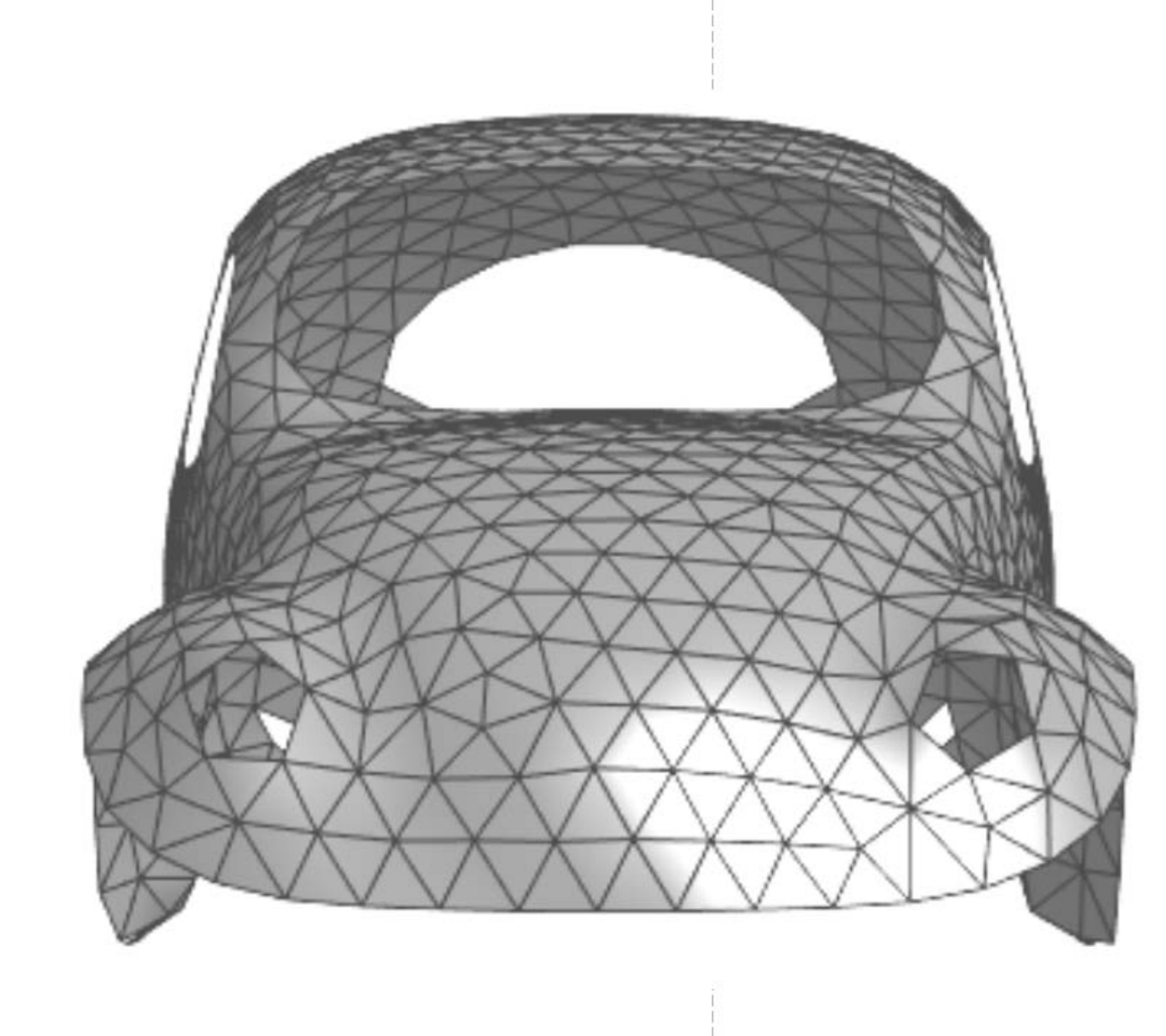}

    \includegraphics[width=0.17\textwidth]{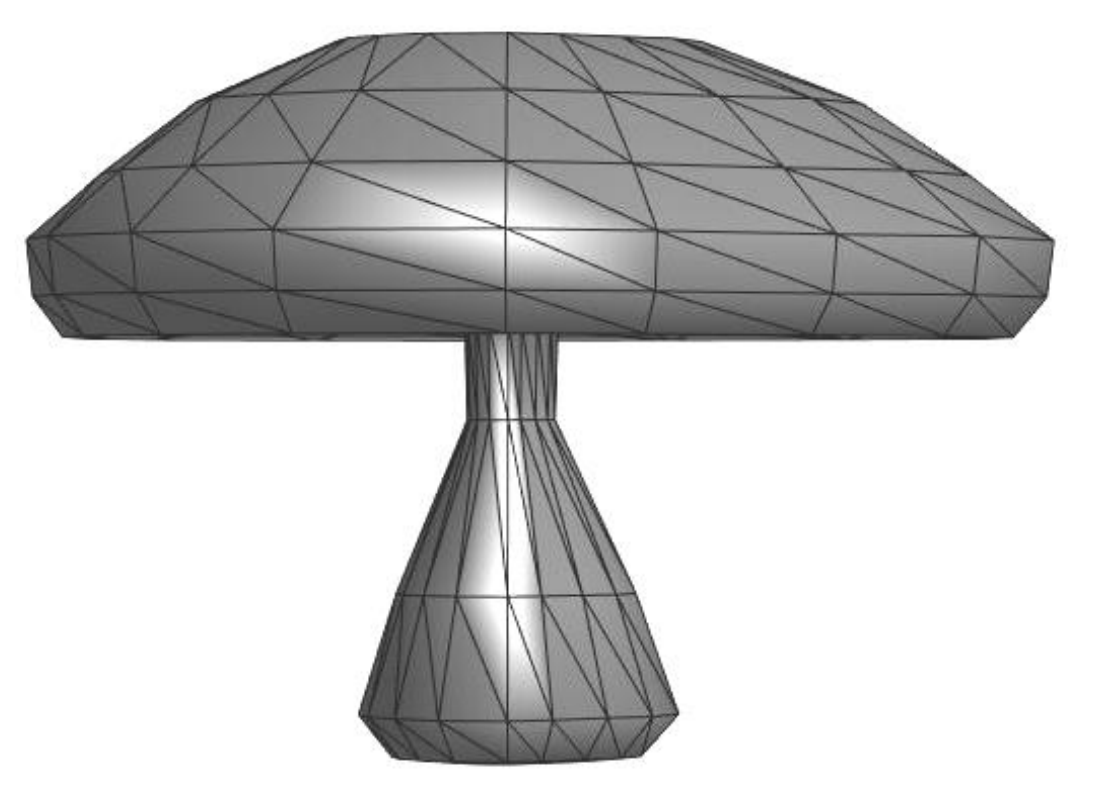}
    \includegraphics[width=0.14\textwidth]{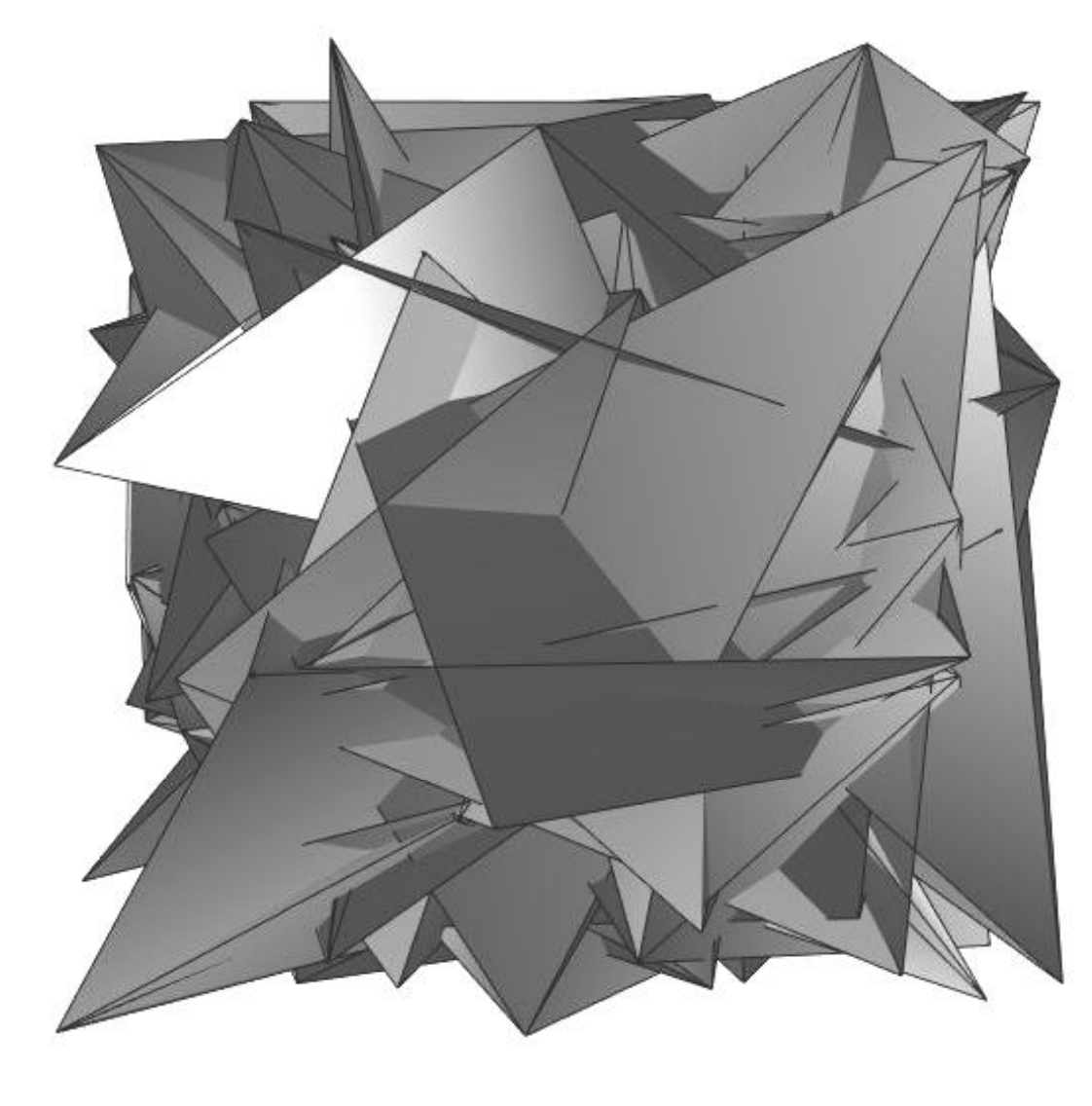}
    \includegraphics[width=0.16\textwidth]{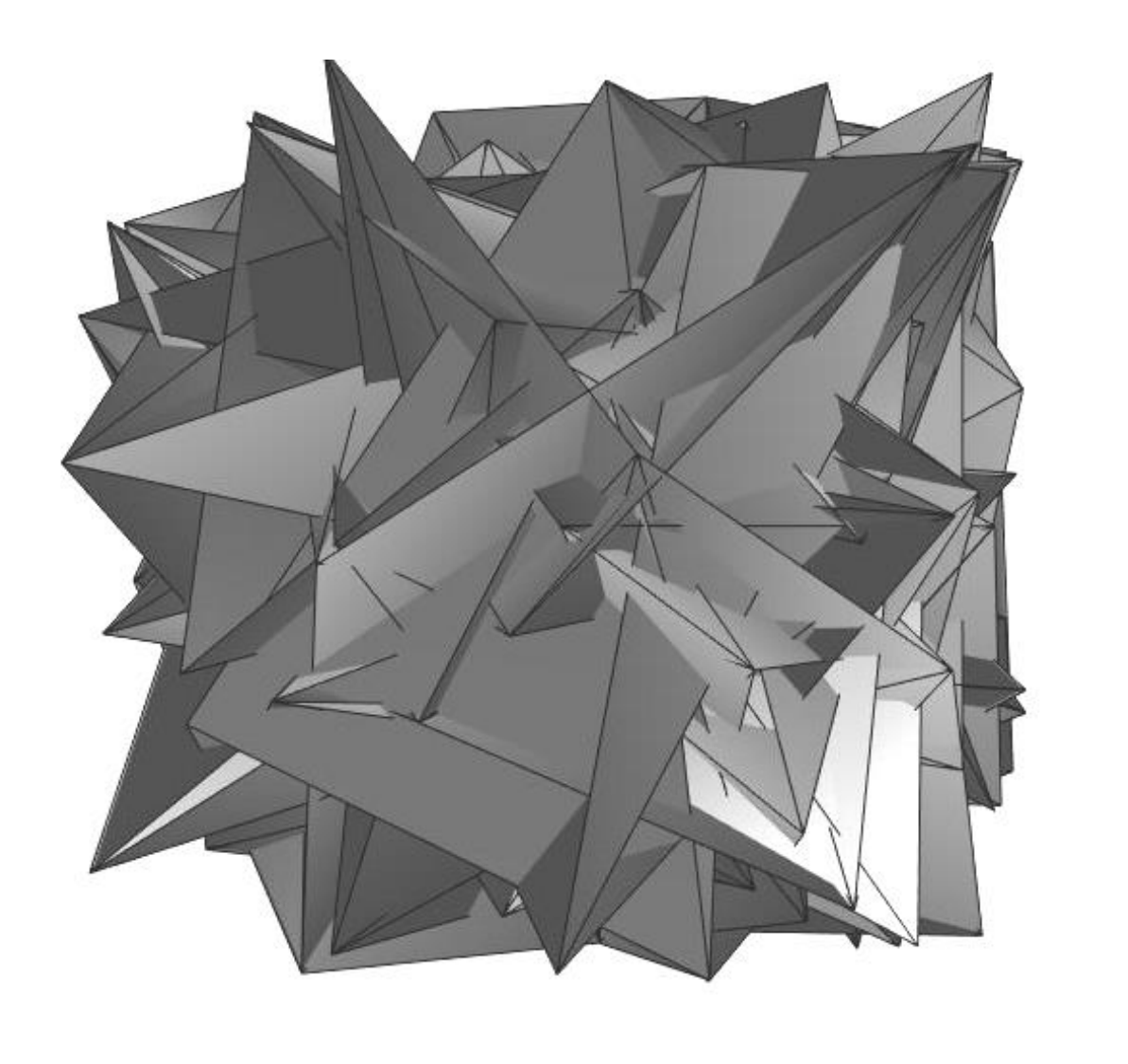}
    \includegraphics[width=0.17\textwidth]{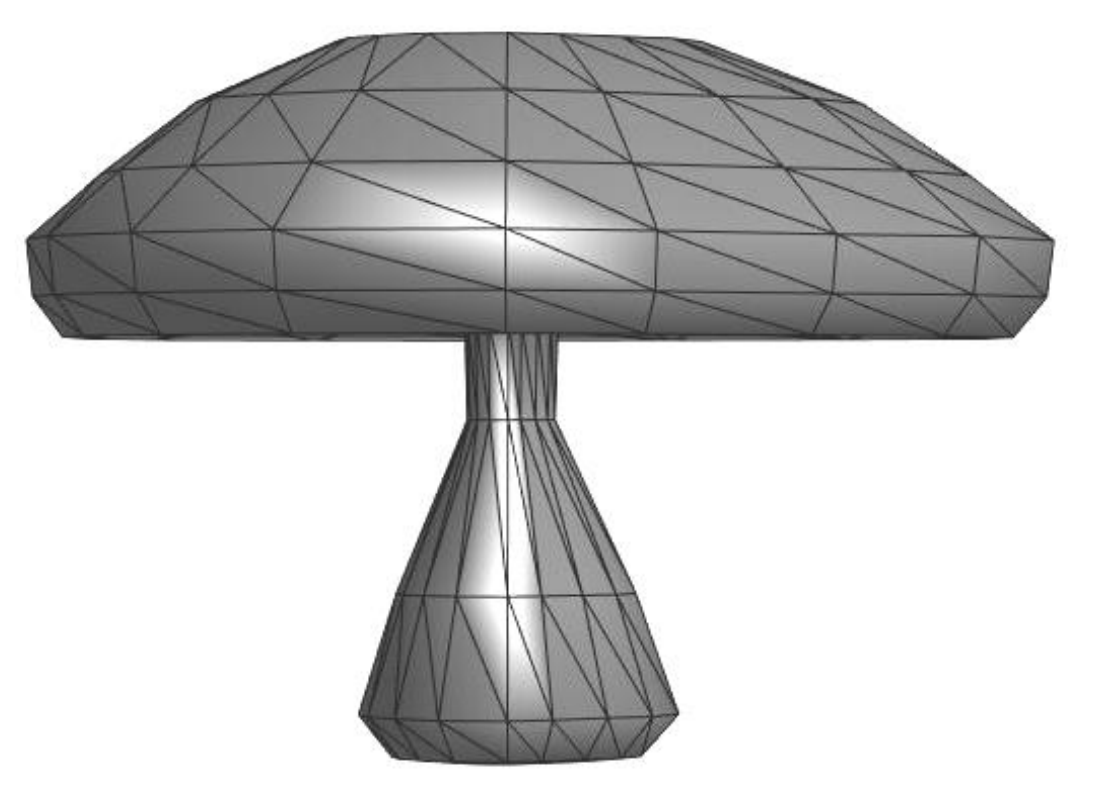}

    \includegraphics[width=0.17\textwidth]{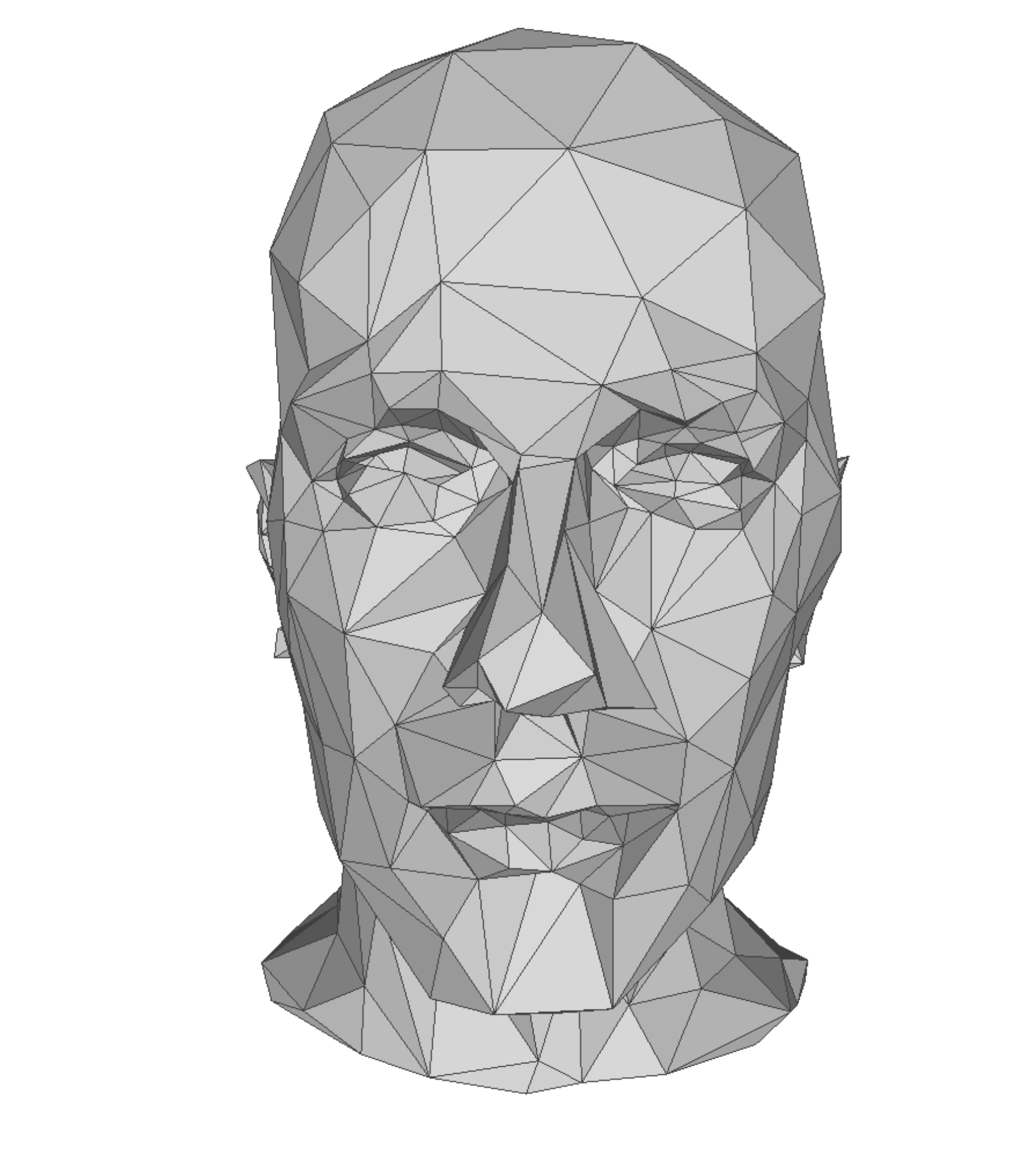}
    \includegraphics[width=0.15\textwidth]{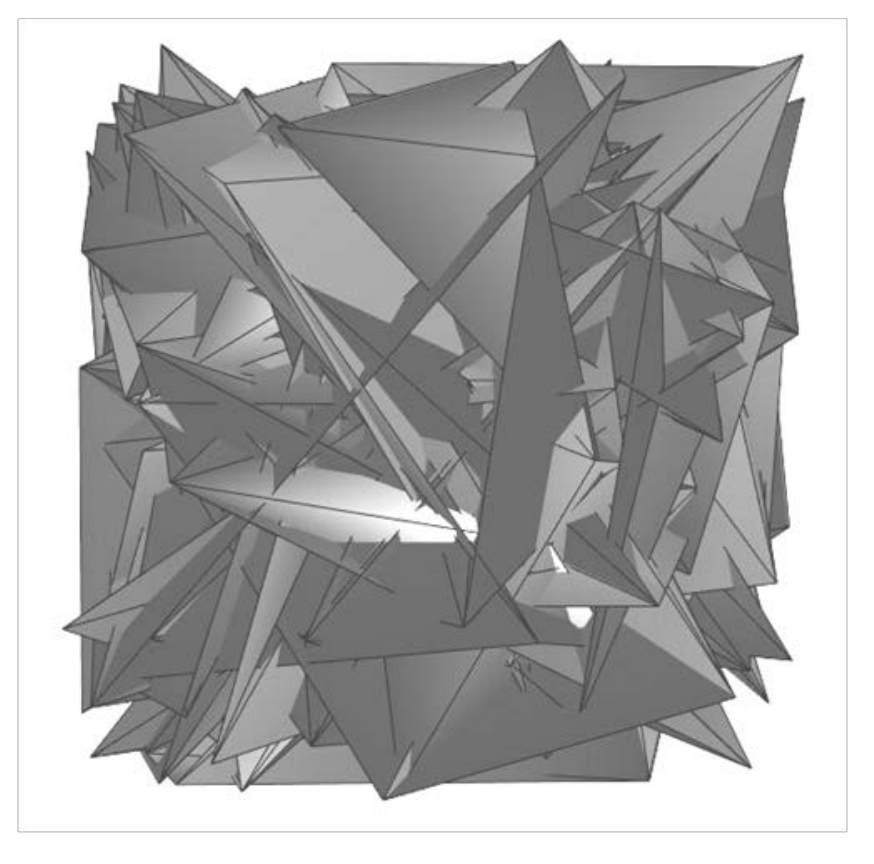}
    \includegraphics[width=0.14\textwidth]{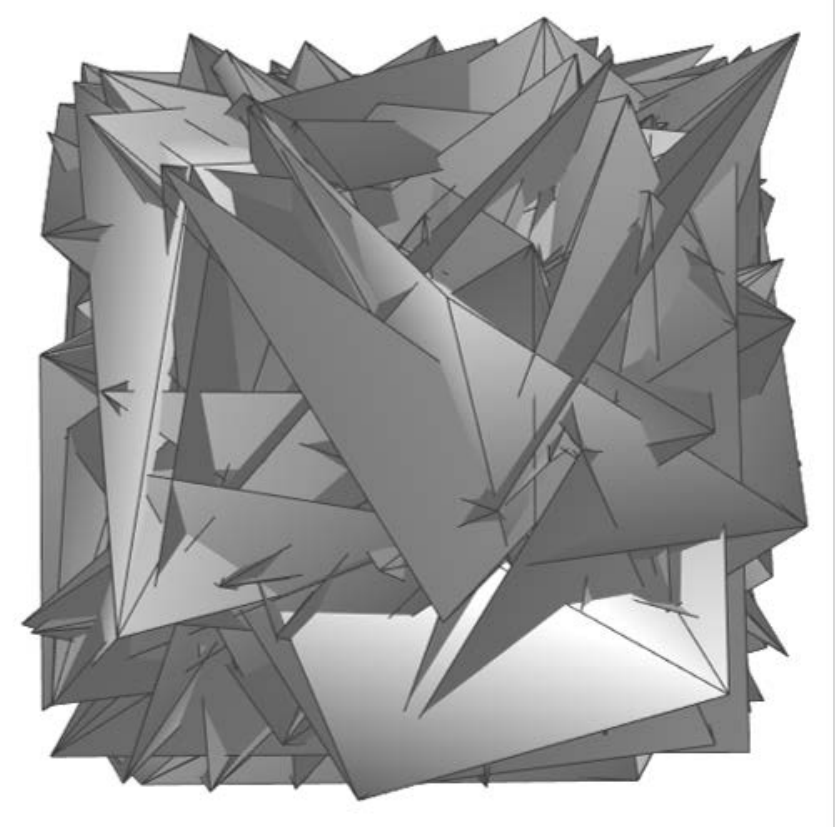}
    \includegraphics[width=0.17\textwidth]{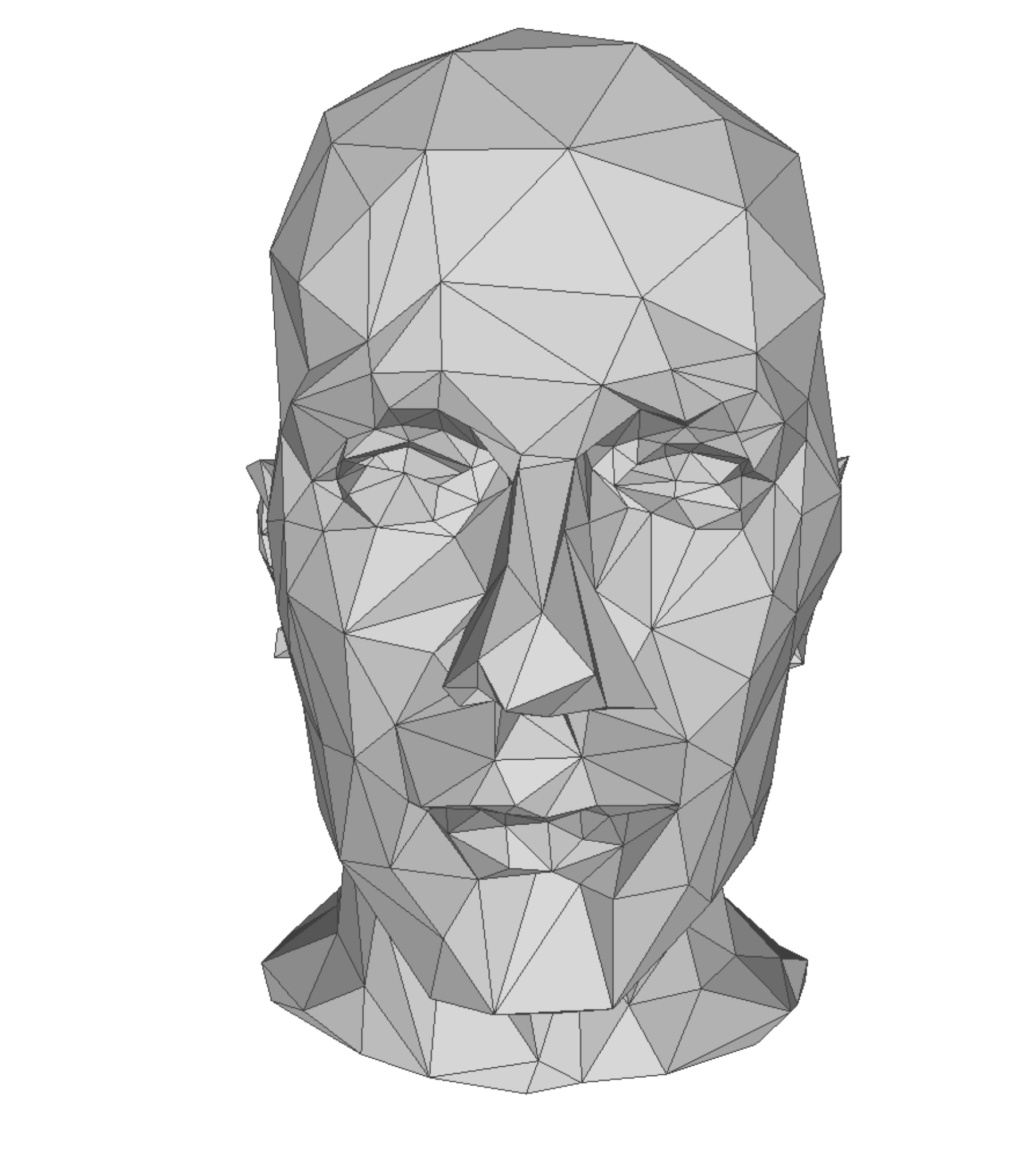}
\label{fig8}
\end{figure*}

\begin{figure*}[!ht]
\vspace{-1cm}  
  \centering
     \subfigure[]{
   \label{fig-9-a}
    \includegraphics[width=0.14\textwidth]{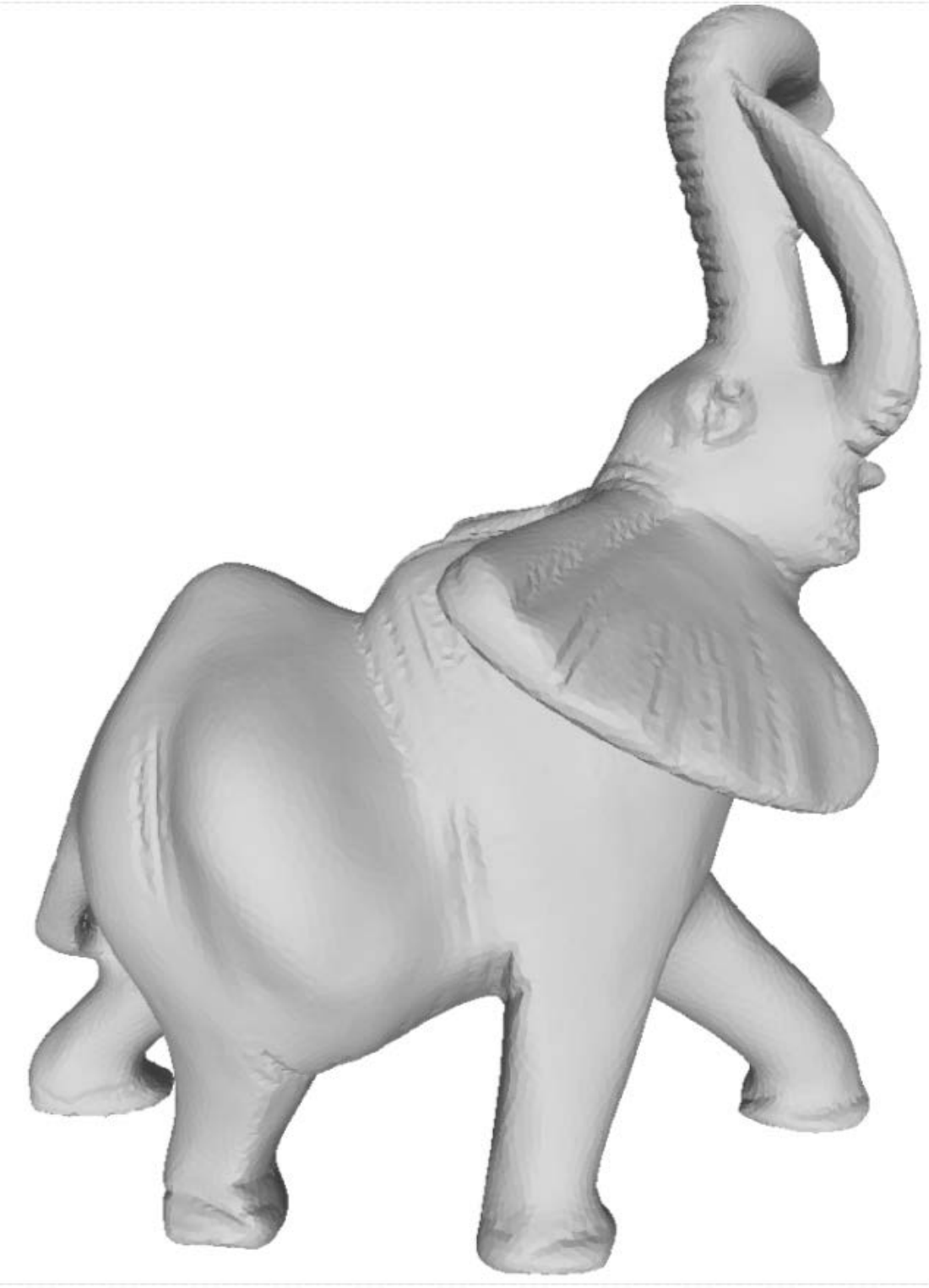}
  }
   \subfigure[]{
   \label{fig-9-c}
    \includegraphics[width=0.14\textwidth]{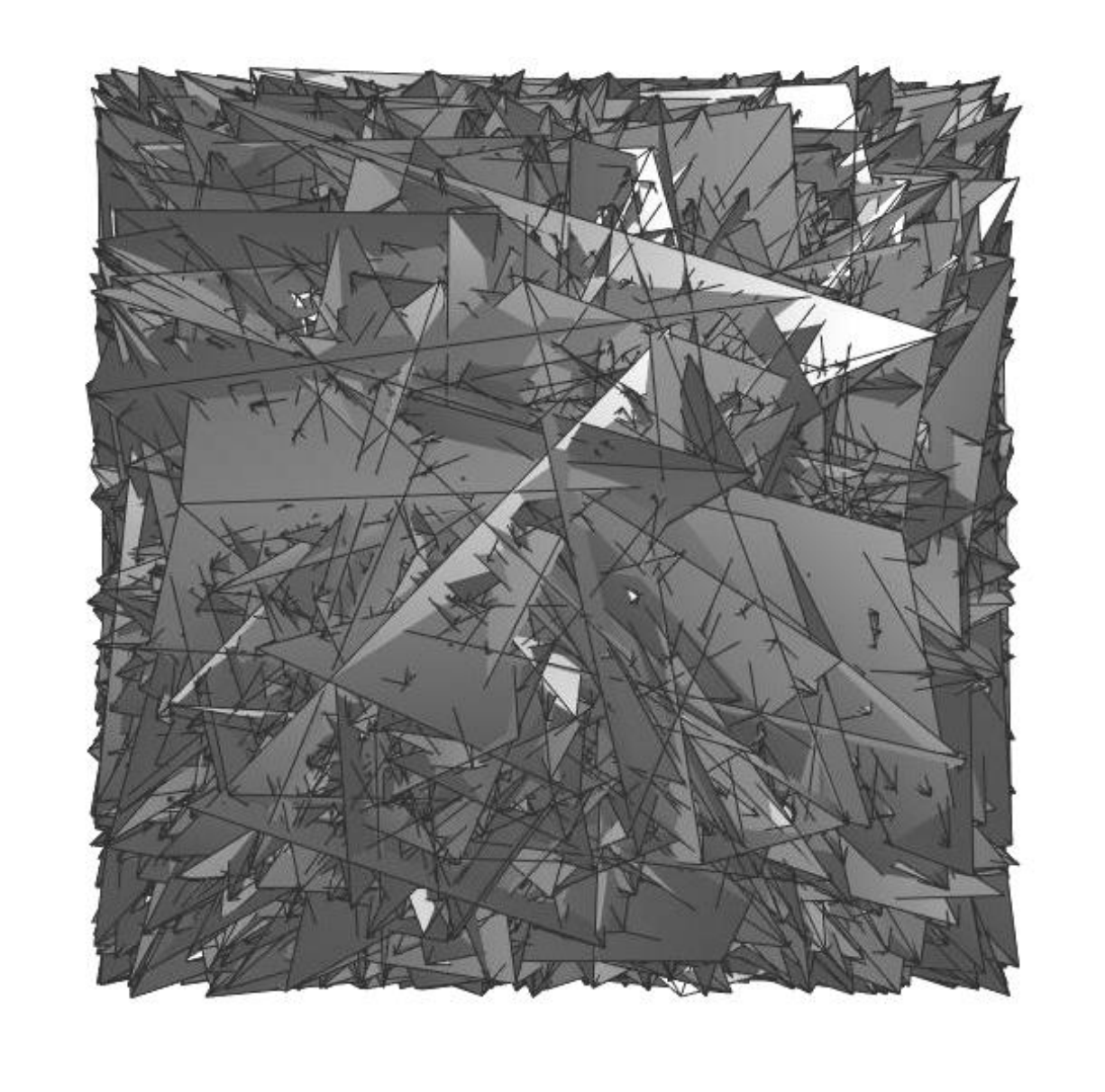}
  }
   \subfigure[]{
   \label{fig-9-b}
    \includegraphics[width=0.14\textwidth]{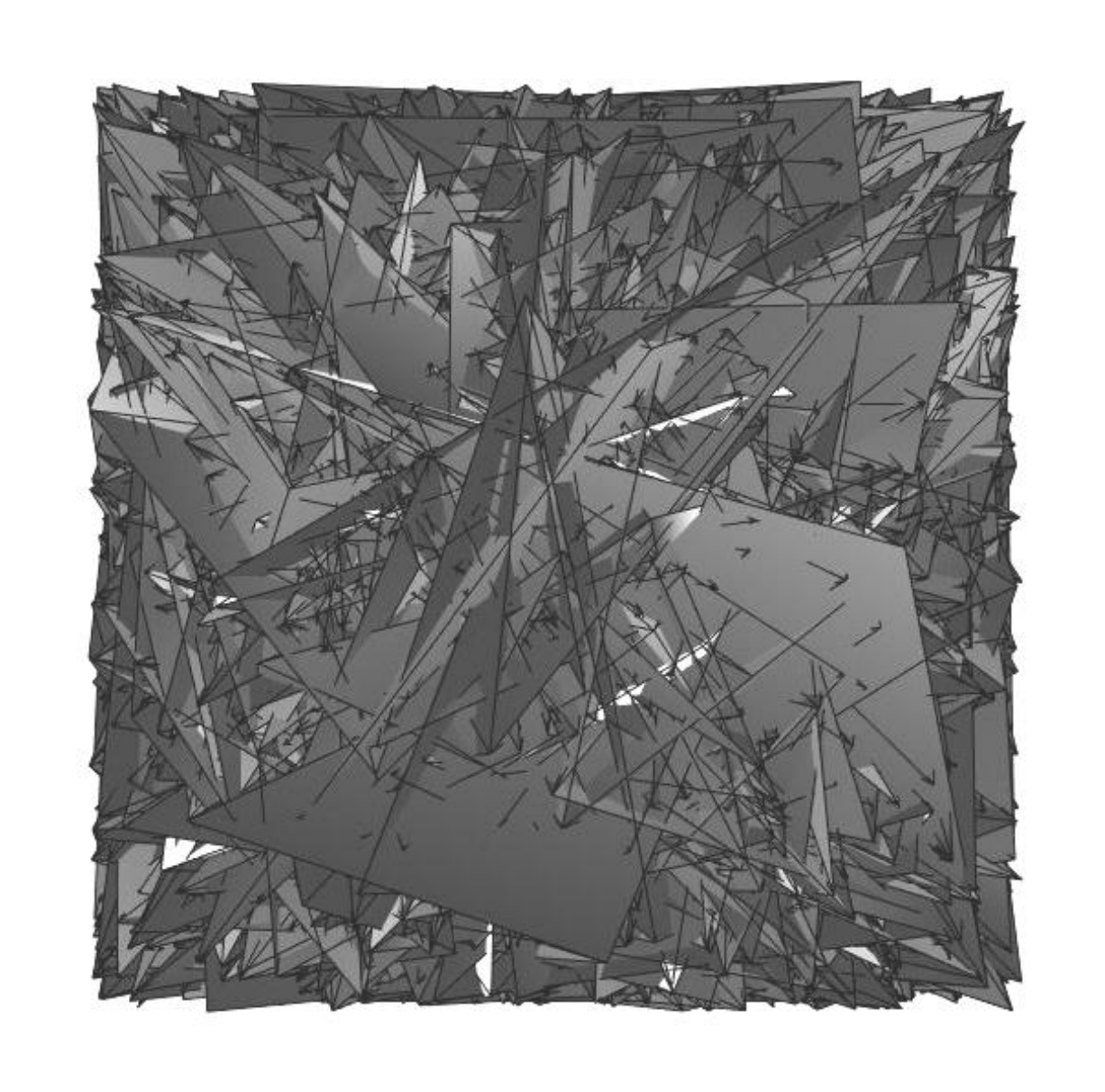}

  }
   \subfigure[]{
   \label{fig-9-d}
    \includegraphics[width=0.14\textwidth]{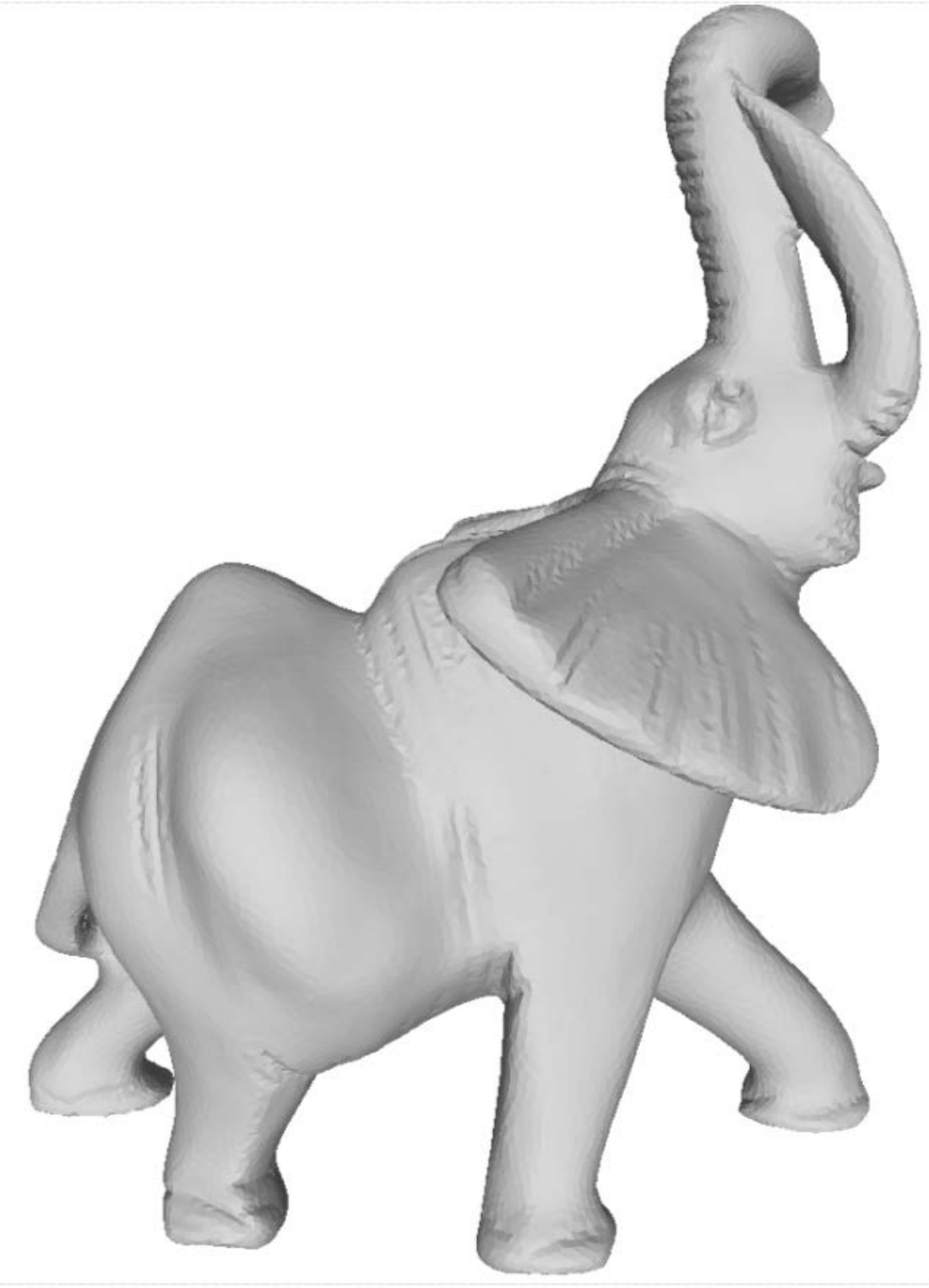}
  }
  \caption{ Illustrative examples showing the appearance of the mesh of each phase : (a)Original mesh, (b)Encrypted mesh, (c) Marked encrypted mesh, (d)Recovered mesh.}
\label{fig9}
\end{figure*}
We performed experiments aimed at dense meshes consisting of tens of millions or hundreds of millions of triangles, which are widely used in 3D printing. We randomly selected three dense meshes in .ply consisting of tens of millions or hundreds of millions of triangle from the The Stanford 3D Scanning Repository to test the performance of our method, which are shown in Fig. 9. Experimental results of dense meshes listed in Table IV illustrate that the embedding rate is fairly high and the data extraction error rate is zero. The experiments shows the applicability and effectiveness of the proposed method to dense meshes.

\begin{table*}[!ht]
\begin{center}
\newcommand{\tabincell}[2]{\begin{tabular}{@{}#1@{}}#2\end{tabular}}
\caption{\label{tb::tab3}Performance of reversible data hiding in dense meshes..}
\setlength{\tabcolsep}{0.7mm}
\begin{tabular}{@{}cccccccc@{}}
\toprule
 \renewcommand\arraystretch{1.3}
Meshes & \tabincell{l}{Number of \\ vertices}  & \tabincell{l}{Number of \\faces}& \tabincell{l}{Embedding \\Rate (bpv)}&\tabincell{l}
{Embedding\\ Capacity (bit)} &\tabincell{l}{Hausdorff \\distance \emph{${10}$$^\emph{$\emph{-3}$}$}}& SNR  & \tabincell{l}{Data error \\percent} \\ 
\midrule
Dragon &  871414 & 437645 & 17.3890& 15153018& 0.0086&101.1375&0$\%$\\
Armadillo & 172974  & 345944 &14.8133&2562315 & 0.0004&104.1417&0$\%$\\
Happyvrip &  543652  & 1087716 & 19.6532&10684501& 0.0008& 100.4581&0$\%$\\
 \bottomrule
\end{tabular}
\end{center}
\end{table*}
\vspace{-0.6cm}
\section{CONCLUSIONS}
In this paper, We proposed an method of RDH-ED using integer mapping and Multi-MSB prediction for encrypted 3D mesh models. Proposed method highlights not only feasible and efficient RDH-ED in 3D meshes also a balance between capacity and distortion.
The Multi-MSB of ``embedded" vertex is replaced by additional data.
Then, due to the spatial correlation of the original mesh, recipient can perfectly predict the Multi-MSB of the ``embedded" vertex by the Multi-MSB of the adjacent vertices around the ``embedded" vertex. 
The Multi-MSB embedding strategy was adopted to improve the embedding capacity, and high quality recovered mesh can be obtained by using ring-prediction in the recovery stage.
The recipient can use the data hiding key to extract the data from encrypted domain and use the encryption key to recover the original mesh separately. 
Experiments show that our method has higher embedding capacity and higher quality recovery mesh compared with the state-of-the-art methods. 

For future work, as the selection of the ``embedded" set is limited by the connectivity of the mesh, it's embedding capacity is not very ideal. We will design a more effective method for selecting the ``embedded" set to improve embedding capacity.

\begin{figure*}[!ht]
\vspace{-0.6cm}
\setlength{\belowcaptionskip}{-0.2cm} 
\setlength{\abovecaptionskip}{-0.2cm} 
  \centering
     \subfigure[]{
   \label{fig-4-a}
    \includegraphics[width=0.21\textwidth]{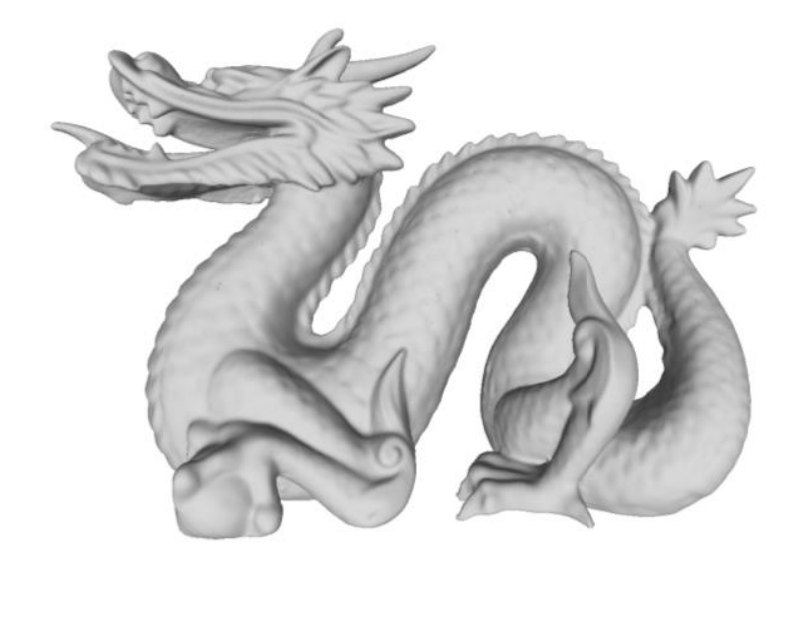}
  }
   \subfigure[]{
   \label{fig-4-c}
    \includegraphics[width=0.22\textwidth]{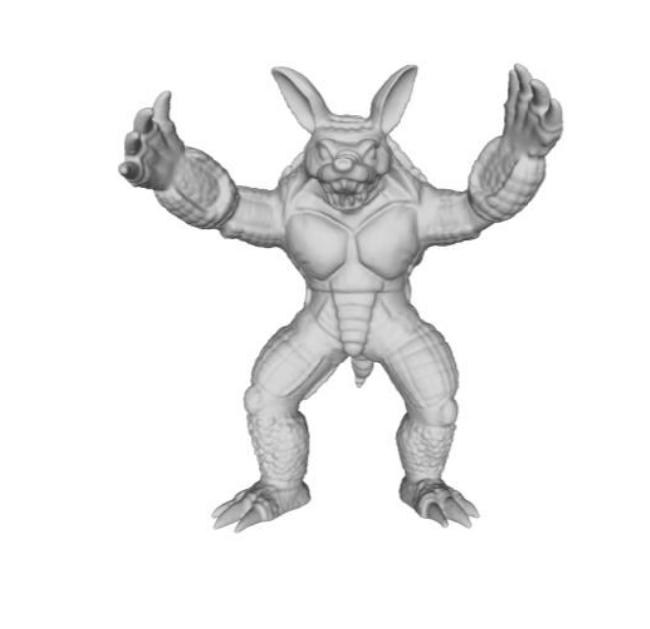}
  }
   \subfigure[]{
   \label{fig-4-b}
    \includegraphics[width=0.22\textwidth]{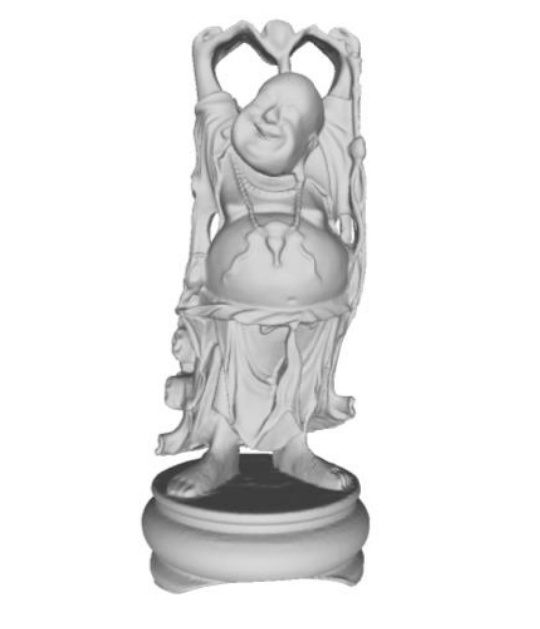}

  }
  \caption{ Dense meshes: (a) Dragon, (b) Armadillo, (c) Happyvrip.}
\label{fig4}
\end{figure*}
\section*{Acknowledgment}
This research work is partly supported by National Natural Science Foundation of China (61872003, 61860206004).

\ifCLASSOPTIONcaptionsoff
  \newpage
\fi


%

\end{document}